\documentclass[letterpaper,titlepage,11pt]{article}
\usepackage{hyperref}

\usepackage{amssymb,amsmath,amsfonts}
\usepackage{epsfig}

\def\clock{{\count0=\time
           dvide\count0 60
           \ifnum\count0<10 0\fi\the\count0
           \multiply\count0 -60 \advance\count0 \time
           :\ifnum\count0<10 0\fi \the\count0
         }}
\newcommand{\timestamp}{{\small\vbox{\hbox{\tt\jobname.tex}
\hbox{\the\day/\the\month/\the\year, \clock}}}}

\def\BB{{\cal B}}
\def\CC{{\cal C}}

\def\EE{{\cal E}}

\def\HH{{\cal H}}

\def\KK{{\cal K}}
\def\LL{{\cal L}}
\def\MM{{\cal M}}

\def\OO{{\cal O}}

\def\SS{{\cal S}}
\def\TT{{\cal T}}

\def\WW{{\cal W}}

\def\d{{\partial}}

\newcommand{\R}{\mathbb{R}}
\newcommand{\T}{\mathbb{T}}
\newcommand{\N}{\mathbb{N}}

\newcommand{\spa}{\ , \ \ }

\newcommand{\beq}{\begin{equation}}
\newcommand{\eeq}{\end{equation}}
\newcommand{\ben}{\begin{displaymath}}
\newcommand{\een}{\end{displaymath}}
\newcommand{\beqa}{\begin{eqnarray}}
\newcommand{\eeqa}{\end{eqnarray}}
\newcommand{\bea}{\begin{eqnarray}}
\newcommand{\eea}{\end{eqnarray}}
\newcommand{\bean}{\begin{eqnarray*}}
\newcommand{\eean}{\end{eqnarray*}}
\newcommand{\ba}{\begin{array}}
\newcommand{\ea}{\end{array}}
\newcommand{\bi}{\begin{itemize}}
\newcommand{\ei}{\end{itemize}}
\newcommand{\ie}{{\it i.e.,\,}}
\newcommand{\eg}{{\it e.g.,\,}}

\newcommand{\lsim}{\mathrel{\raisebox{-.8ex}{$\stackrel{\textstyle<}{\sim}$}}}

\newcommand{\vep}{\varepsilon}

\numberwithin{equation}{section}

\begin{document}

\begin{titlepage}
\begin{flushright}
NORDITA-2009-78\\
CPHT-RR135.1209\\
\end{flushright}
\vskip 2.5cm
\begin{center}
{\bf\LARGE New Horizons for Black Holes and Branes}
\vskip 1.6cm
{\bf Roberto Emparan$^{a,b}$, Troels Harmark$^{c}$, Vasilis Niarchos$^{d}$,}
 {\bf Niels A. Obers$^{e}$ } \vskip 0.5cm
\medskip
\textit{$^{a}$Instituci\'o Catalana de Recerca i Estudis
Avan\c cats (ICREA)}\\
\smallskip
\textit{$^{b}$Departament de F{\'\i}sica Fonamental and}\\
\textit{Institut de Ci\`encies del Cosmos, Universitat de Barcelona,}\\
\textit{Mart\'{\i} i Franqu\`es 1, E-08028 Barcelona, Spain}\\
\smallskip
\textit{$^{c}$NORDITA, Roslagstullsbacken 23, SE-106 91 Stockholm,
Sweden}\\
\smallskip
\textit{$^{d}$Centre de Physique Th\'eorique, \'Ecole Polytechnique,
 91128 Palaiseau, France}\\
 \textit{Unit\'e mixte de Recherche 7644, CNRS}\\
\smallskip
\textit{$^{e}$The Niels Bohr Institute,} \textit{Blegdamsvej 17,
2100 Copenhagen \O, Denmark}

\vskip .2 in
{\tt emparan@ub.edu, harmark@nordita.org, niarchos@cpht.polytechnique.fr,
obers@nbi.dk}
\end{center}
\vskip 0.1in

\baselineskip 16pt
\date{}

\begin{center}
{\bf Abstract}
\end{center}
\vskip 0.2cm

\noindent We initiate a systematic scan of the landscape of black holes
in any spacetime dimension using the recently proposed blackfold
effective worldvolume theory. We focus primarily on asymptotically flat
stationary vacuum solutions, where we uncover large classes of new black
holes. These include helical black strings and black rings, black
odd-spheres, for which the horizon is a product of a large and a small
sphere, and non-uniform black cylinders. More exotic possibilities are
also outlined. The blackfold description recovers correctly the
ultraspinning Myers-Perry black holes as ellipsoidal even-ball
configurations where the velocity field approaches the speed of light at
the boundary of the ball. Helical black ring solutions provide the first
instance of asymptotically flat black holes in more than four dimensions
with a single spatial $U(1)$ isometry. They also imply infinite rational
non-uniqueness in ultraspinning regimes, where they maximize the entropy
among all stationary single-horizon solutions. Moreover, static blackfolds are
possible with the geometry of minimal surfaces. The absence of compact
embedded minimal surfaces in Euclidean space is consistent with the
uniqueness theorem of static black holes.

\end{titlepage} \vfill\eject

\setcounter{equation}{0}

\pagestyle{empty}
\small
\tableofcontents
\normalsize
\pagestyle{plain}
\setcounter{page}{1}

\newpage

\section{Introduction}
\label{intro}

Recently we presented the general principles of a new effective
description of higher-dimensional black holes that captures the
long-distance physics of black holes with horizons that possess at least
two widely separate length-scales \cite{Emparan:2009cs,Emparan:2009at}.
In such regimes of parameter space the black hole is regarded as a black
brane curved into a submanifold of a background spacetime ---a {\it
blackfold}--- whose dynamics can be formulated in terms of an effective
fluid that lives on a dynamical worldvolume. It is expected that the
equations of motion of this effective worldvolume theory guarantee that
the corresponding black hole solution is regular on and outside the
horizon. These equations split into {\it intrinsic} (fluid-dynamical)
and {\it extrinsic} (generalized geodesic embedding) equations.

The results in \cite{Emparan:2009cs,Emparan:2009at,Emparan:2007wm}
show that the blackfold approach does indeed reproduce in a precise manner
\textit{all} the long-distance physics of known black hole and black
brane solutions in $D\geq 5$ vacuum Einstein gravity, namely: the
existence of thin black rings in $D=5$; the ultraspinning regimes of
Myers-Perry (MP) black holes in $D\geq 6$; and the long-wavelength
component of the Gregory-Laflamme instability of black branes. Moreover,
the derivation of all these phenomena in the blackfold approach is
remarkably simpler than in an analysis of the full Einstein
equations. Thus the method appears to be a highly efficient tool to
investigate black hole physics possibly lying well beyond the
conceivable reach of exact techniques.

In this work we explore specific solutions of the blackfold equations
aiming to uncover \textit{qualitatively new} features of black holes and
branes in higher dimensional gravity ---new horizon topologies, new
types of stationary solutions with a minimal set of isometries and new
examples of non-uniqueness. The results, which allow us to probe special
corners of the phase diagram of higher-dimensional black holes, provide
useful input towards a more global understanding of the full phase
diagram, which is known to exhibit a rich pattern of interconnected
phases and merger points with topology changing transitions
\cite{Emparan:2007wm}%
\footnote{See \cite{Obers:2008pj,Niarchos:2008jc}
for brief reviews of higher-dimensional black holes and
\cite{Emparan:2008eg} for a more extensive one.}.

As a first step we apply the formalism to the simplest,
universal case of asymptotically flat stationary black hole
solutions of pure Einstein gravity in $D$ dimensions. In this case,
the asymptotic charges that characterize a black hole are the mass
$M$ and $\left\lfloor \frac{D-1}{2} \right\rfloor$ angular momenta $J_i$.
Blackfolds are an effective description of black holes in the
ultraspinning regime where $\ell_J \gg \ell_M$, with $\ell_M \sim
(GM)^{\frac{1}{D-3}}$ and $\ell_J\sim \frac{J}{M}$ the
characteristic scales of the problem.\footnote{$J=\Big(
\displaystyle{\sum_i J_i^2}\Big)^{1/2}$ aggregates the effect of all
possible angular momenta.} In this paper we work, for the most part,
to leading order in the expansion at small $\ell_M/\ell_J$ in which the
black brane is treated at probe level, but we also discuss their
backreaction in some
particular cases.

The blackfold methodology has been applied already to black rings in
(A)dS spaces \cite{Caldarelli:2008pz}, to black rings in Taub-NUT spaces
\cite{Camps:2008hb}, to black strings in plane waves in \cite{LeWitt:2009qx},
and to the simplest dipole-charged and supersymmetric black rings
\cite{BlancoPillado:2007iz,Caldarelli:2008pz}. More general charged
solutions (\eg\ in supergravity theories) or in curved backgrounds (\eg\
in the presence of a cosmological constant), the extension to
time-dependent non-stationary solutions, and related issues of
stability, will be treated extensively in forthcoming publications.

We now outline the structure of the paper and highlight the main results.
We begin in Section \ref{equations} by briefly reviewing the key
points of the blackfold formalism. We then present the relevant
equations for stationary neutral blackfolds including an action
principle. We also discuss how to analyse the physical properties of
blackfolds. The main results of this paper are in Sections
\ref{1folds} to \ref{static} where we present various stationary
neutral blackfold configurations. We focus mainly on asymptotically
flat solutions in a Minkowski background.

We begin in Section \ref{1folds} by considering general stationary
configurations of black one-folds. A black one-fold is a black
string bent on a spatial curve. This includes the black ring as a
special case, but we find that there are solutions other than the black ring.
In fact we can find the most general black one-fold in a Minkowski
background, and show that generically it traces a helicoidal curve.
{\it Helical black rings} exhibit a number of remarkable properties,
such as an infinite non-uniqueness labelled by rational parameters
and maximal entropy among single-horizon black holes in
ultraspinning regimes with at least two ultraspins. But perhaps their
most striking property is that generically they preserve only two commuting
Killing vector fields ---one timelike, another spacelike--- among the
$\left\lfloor\frac{D+1}{2}\right\rfloor$ possible commuting symmetries
of an asymptotically flat stationary solution. Two
is the minimal set of commuting Killing vectors allowed by general theorems
\cite{Hollands:2006rj, Moncrief:2008mr}, and it has been conjectured that
there should exist stationary, asymptotically flat black hole solutions of
the vacuum Einstein equations that admit exactly
two commuting Killing vector fields \cite{Reall:2002bh}. Helical
black rings therefore constitute the first example of such solutions in
every dimension $D\geq 5$.

We continue in Section \ref{oddsphere} by generalizing
the black ring to solutions with horizon topology $\prod S^{p_a} \times s^{n+1} $
(with the product over odd $p_a$) rotating rigidly along the $\frac{p_a+1}{2}$ Killing
isometries of each sphere. We will mostly consider solutions with
{\it round} odd-spheres, but more general situations can be
envisioned (a concrete example of a non-round three-sphere is discussed in
appendix \ref{s3}). Besides the black ring this family includes a plethora of
novel black hole solutions, including examples like black odd-spheres,
black tori, \textit{etc}. For the case of black tori
with horizon topology $\T^p \times s^{n+1}$, appendix \ref{tori} discusses
the leading-order perturbative solution of the metric, generalizing
the corresponding result for black rings obtained in Ref.~\cite{Emparan:2007wm}.

The blackfold equations do not have analogous solutions with
even-spheres, but they do admit ellipsoidal even-ball configurations
where the velocity field approaches the speed of light at the boundary
of the ball. Interestingly, these configurations, which are examined in
Section \ref{ball}, describe black holes with spherical horizon topology
and capture faithfully the properties of ultraspinning Myers-Perry black
holes.

In Section \ref{nubc} we present a one-parameter family of
inhomogeneous black cylinder configurations. The axisymmetric
inhomogeneity resembles the Rayleigh-Plateau inhomogeneity of
cylindrical streams of fluid in hydrodynamics. One can solve the
blackfold equations for generic values of the inhomogeneity
parameter until the onset of a non-linear regime where the solution
develops short necks and we exit the regime of validity of our
approximations. We also discuss the case when one of the directions of the
background is compactified and the cylindrical blackfold is wrapped
around this direction. The resulting phase diagram bears strong
resemblance to that of black strings and holes
in Kaluza-Klein space and suggests a horizon-topology changing phase
transition analogous to the one observed in Ref.~\cite{Kudoh:2004hs}
(see \cite{Kol:2004ww,Harmark:2007md} for reviews).

Finally, in Section \ref{static} we briefly consider static black
brane configurations. In this case, the generic solution of the
blackfold equations is a minimal hypersurface ($i.e.$ a hypersurface
with vanishing mean curvature vector). One can use the rich
mathematical literature  on minimal surfaces (see for example \cite{Colding})
to obtain corresponding
{\it minimal} blackfolds. However, compact embedded minimal
surfaces in Euclidean space do not exist. We point out how
the blackfold construction relates this mathematical result
to the uniqueness theorem for static black holes
in vacuum gravity \cite{Gibbons:2002bh}.

In Section \ref{phasescan} we summarize the emerging phases of
higher-dimensional black holes and compare their entropy. We also
outline in Section \ref{multibhs} how configurations with multiple black
holes can be described using the blackfold approach, and disprove a
conjecture made in \cite{Emparan:2007wm} about the existence of
`pancaked black Saturns' in thermodynamic equilibrium. We conclude in
Section \ref{conclusions} with a more general discussion of the
philosophy of the blackfold approach in a broader context, and the
possibility of explicitly constructing new metrics perturbatively. We
also discuss interesting open issues for further research, in particular
those related to the full phase diagram of higher-dimensional black
holes, dynamical aspects and further new solutions.

Given the background material in Sec.\ 2 each of the Sections 3--7 is
self-contained and can be read independently.

Note that throughout the paper we use the notation that $D$ is the
spacetime dimension, $p$ is the spatial dimension of the blackfold
worldvolume and $n$ is defined by
\beq \label{notaa}
n=D-p-3\geq 1\,.
\eeq
The volume of the $n$-sphere is denoted $\Omega_{(n)}$ and the angular
velocities $\Omega_i$. The volume of spatial sections of the blackfold
is $V_{(p)}$, the velocity field is $V$ and its components $V^i$.

\section{Blackfold equations}
\label{equations}

In the blackfold approach the black hole is
described effectively as a thin black $p$-brane curved on a
submanifold $\mathcal{W}_{p+1}$ embedded in the background
space-time. The degrees of freedom associated with the scale of the
thickness of the $p$-brane are integrated out and are described
effectively in terms of a stress tensor $T_{\mu\nu}$ that is
supported on $\mathcal{W}_{p+1}$. The submanifold
$\mathcal{W}_{p+1}$ is characterized by an embedding $X^\mu(\sigma)$
that depends on the worldvolume coordinates $\sigma^a$, $a=0,1...,p$,
and with $\mu = 0,1,...,D-1$. From this one can obtain the induced
worldvolume metric $\gamma_{ab} =
\partial_a X^\mu \partial_b X^\nu g_{\mu\nu}$, with $g_{\mu\nu}$
being the metric of the background space-time, and the first
fundamental form $h^{\mu\nu} =
\partial_a X^\mu
\partial_b X^\nu \gamma^{ab}$, which acts as a projector tangential
to the submanifold $\mathcal{W}_{p+1}$. This can furthermore be used
to define $\perp_{\mu\nu} = g_{\mu\nu} - h_{\mu\nu}$, which projects
in directions orthogonal to $\mathcal{W}_{p+1}$. Finally, the shape of
the embedding is characterized by
the extrinsic curvature tensor ${K_{\mu\nu}}^\rho = {h_\mu}^\lambda
{h_\nu}^\sigma \nabla_\sigma {h_\lambda}^\rho$, where
$\nabla_\sigma$ is the covariant derivative in the background spacetime.
From this we obtain  the mean
curvature vector $K^\rho = h^{\mu\nu} {K_{\mu\nu}}^\rho$, which can be obtained as
$K^\rho=\Box X^\rho+h^{\mu\nu}\Gamma^\rho_{\mu\nu}$, where $\Box$ is the
d'Alembertian in the worldvolume metric $\gamma_{ab}$.

The blackfold approach can be used in the regime in which the thickness
of the brane, here denoted $r_0(\sigma)$, is much smaller than the
characteristic scale of the geometry of the submanifold
$\mathcal{W}_{p+1}$, typically set by its extrinsic curvature. The
thickness $r_0(\sigma)$ can be defined in a local rest frame of the flat
static $p$-brane as the
horizon radius of the sphere transverse to the worldvolume. It
determines the energy density and
pressure in the effective stress tensor $T_{\mu\nu}$.

The blackfold equations consist of $D$ equations, resulting from
stress-energy conservation, on an equal number of worldvolume field
variables -- the thickness $r_0
(\sigma)$, the velocity field $u_a(\sigma)$ and the transverse embedding
coordinates $X^\perp (\sigma)$. It was shown in
Ref.~\cite{Emparan:2009at} that the $D$ blackfold equations can be
split up in ({\em a}) a part associated with the intrinsic dynamics on the
brane, which takes the form of fluid dynamics on the brane, and ({\em b}) the
extrinsic
dynamics of the brane, viewing it as a source of energy-momentum
localized on the submanifold $\mathcal{W}_{p+1}$ in the target
space-time. The intrinsic dynamics on the brane is governed by
energy-momentum conservation on the worldvolume, corresponding to
$p+1$ equations, while the extrinsic dynamics is described by Carter's
equation \cite{Carter:2000wv}
\begin{equation}
\label{Carter} T^{\mu\nu} {K_{\mu\nu}}^\rho = 0
\end{equation}
corresponding to $D-p-1$ equations\footnote{The
conservation of the quasilocal
stress energy tensor as the balance condition for five-dimensional
black rings is also discussed in \cite{Astefanesei:2009mc}. However, in
that instance the
quasilocal stress tensor is computed on a surface of topology $S^3$,
instead of $S^1\times S^2$, so the sphere $S^2$ cannot be `integrated out' to
obtain a one-dimensional effective stress tensor
and therefore the analysis of \cite{Astefanesei:2009mc} has no bearing
on the blackfold approach.}.
When applying this equation to specific stress tensors, it is very
convenient to use that, for any vectors $t,s$ that are tangent to the
worldvolume, the projected extrinsic curvature is
\beq\label{tsK}
t^\mu s^\nu {K_{\mu\nu}}^\rho={\perp^\rho}_\mu \nabla_t s^\mu
={\perp^\rho}_\mu \nabla_s t^\mu\,.
\eeq

\subsubsection*{Stationary blackfolds}

In this paper we are primarily interested in finding stationary solutions
of the blackfold equations, and mostly in a Minkowski spacetime
background so the resulting black hole is asymptotically flat.
Stationarity implies that the black brane has a Killing horizon
associated to a Killing vector $\mathbf{k}$ with surface gravity
$\kappa$. As shown in detail in ref.~\cite{Emparan:2009at}, this
isometry enables one to solve the intrinsic blackfold equations, leaving
only the extrinsic equations \eqref{Carter} for the embedding. We
assume that this Killing vector can be written in terms
of orthogonal commuting Killing vectors of the background
spacetime\footnote{Since we only consider spatial Killing vectors, and
combinations thereof, with closed orbits (otherwise the blackfold
approximation presumably breaks down), the rigidity theorem of
\cite{Hollands:2006rj, Moncrief:2008mr} only requires the existence of one spatial
Killing vector. The motivation to write it as a linear combination of
other spatial vectors is to relate it to asymptotic symmetries of the
background.}
\begin{equation}\label{kxichi}
\mathbf{k} = \xi + \sum_i \Omega_i \chi_i
\end{equation}
where $\xi$ is the generator of
time-translations of the background space-time, with canonical unit
normalization at asymptotic infinity, and $\chi_i$ are
generators of angular rotations in the background space-time
normalized such that their orbits have periods $2\pi$. The $\Omega_i$
are thus the corresponding angular velocities. To have a stationary
blackfold the Killing vector fields $\xi$ and $\chi_i$ should
correspond to symmetries of the submanifold $\mathcal{W}_{p+1}$. On the
worldvolume, these vectors will be proportional to a set of orthonormal vectors
$\partial/\partial t$, $\partial/\partial z^i$
\begin{equation}\label{xitchiz}
\xi = R_0(\sigma)\frac{\partial}{\partial t}\,, \qquad \chi_i = R_i(\sigma)
\frac{\partial}{\partial z^i}\,.
\end{equation}
The factor $R_0$ corresponds to the redshift between the worldvolume and
asymptotic infinity. For the most part in this paper we will consider
blackfolds in a
Minkowski background, in which
\beq
R_0=1 \qquad \mathrm{(Minkowski~background)}\,.
\eeq
When more general situations are considered it will be explicitly stated
that $R_0\neq 1$.
$R_i(\sigma)$ are the
proper radii of the orbits of $\chi_i$. With this, the norm of
$\mathbf{k}$ is
\begin{equation}
\label{kV}
 |\mathbf{k}|= \sqrt{1-V^2(\sigma)} \spa V^2 = \sum_i
\Omega_i^2 R_i^2(\sigma)
\end{equation}
where $V$ is interpreted as a velocity field on the blackfold relative
to observers that follow orbits of $\partial/\partial t$. The vector
$\mathbf{k}$ also corresponds to the Killing generator of the
horizon.

Since $\xi$ is a Killing vector field for both the background and
the embedding submanifold $\WW_{p+1}$ it is possible to define the
time-independent spatial section $\BB_p$ of $\WW_{p+1}$.
If in addition the background is static\footnote{Indeed, ultrastatic if
we assume that $\xi$ has constant norm, as will often be the case when
$R_0=1$.} and therefore $\xi$ is
orthogonal to $\BB_p$, the problem of
finding the embedding of $\WW_{p+1}$ in $\R^{1,D-1}$ reduces to
finding an embedding of a spatial $p$-dimensional submanifold
$\BB_p$ in $\R^{D-1}$.

\subsubsection*{Effective stress tensor}

The effective theory of blackfolds is organized as a derivative
expansion in the worldvolume fields, and in this paper we only work to
leading order. This implies that the effective stress tensor has the
perfect fluid form
\begin{equation}
\label{stress1} T^{\mu\nu} = ( \vep + P ) u^\mu u^\nu + P
h^{\mu\nu}
\end{equation}
where for the neutral stationary blackfolds we have
\begin{equation}\label{stress2}
\vep   =\frac{\Omega_{(n+1)}}{16\pi G}(n+1)r_0^n\,,\qquad
P=-\frac{1}{n+1}\,\vep\,,
\end{equation}
with
\beq\label{r0}
r_0  =  \frac{n \sqrt{1-
V^2}}{2\kappa}\,,
\eeq
and
\begin{equation}\label{uvel}
u
=\frac{1}{\sqrt{1-
V^2}}\left(\xi + \sum_i \Omega_i \chi_i\right)=\frac{1}{\sqrt{1-
V^2}}\left(\frac{\partial}{\partial t}+\sum_i V^i\frac{\partial}{\partial z^i}\right)
\,,
\eeq
where $V^i =R_i\Omega_i$ and $\Omega_{(n+1)}$ is the volume of the unit
$(n+1)$-sphere.

\subsubsection*{Blackfold equations and action}

The extrinsic equations for stationary blackfolds are obtained by
inserting the stress tensor \eqref{stress1}-\eqref{stress2} in the
Carter equation \eqref{Carter}. This yields $D-p-1$ equations,
the solutions of which describe stationary blackfolds to leading
order in the `test' brane approximation. Using \eqref{tsK} these equations can
be written more concisely as
\begin{equation}
\label{BFeqs} K^\rho = n \perp^{\rho \mu } \partial_\mu \log \left(
\sqrt{1-V^2} \right)\,,
\end{equation}
which can equivalently be found by varying the action
\begin{equation}
\label{BFact} I = \beta \int_{\BB_p} d V_{(p)} \left(1-V^2\right)^{\frac{n}{2}}\,,
\end{equation}
where $dV_{(p)}$ is the integration measure on $\BB_p$ and the trivial
integration over a time interval $\beta$ has been performed. The
action
\eqref{BFact} is in many applications the most efficient way to derive
specific forms of the stationary blackfold equations \eqref{BFeqs}.

\subsubsection*{Mass, angular momentum and entropy}

For given values of the temperature $T = \kappa / (2\pi)$ and the
angular velocities $\Omega_i$, solving the stationary blackfold
equations amounts to determining the
embedding of the worldvolume, which contains in particular the functions
$R_i(\sigma)$. These in turn determine the intrinsic fields of the
blackfold, namely the
velocity $V(\sigma)$ in \eqref{kV}, and the thickness $r_0(\sigma)$, \ie
the horizon radius transverse to the
worldvolume, in \eqref{r0}.
The mass and angular momenta are obtained by integrating the stress
tensor as
\beq\label{mass} M=\frac{\Omega_{(n+1)}}{16 \pi G}
\left(\frac{n}{2\kappa}\right)^n \int_{\mathcal{B}_p}dV_{(p)}\;
(1-V^2)^\frac{n-2}{2}\left(n+1-V^2\right)\,, \eeq and
\beq\label{angmom} J_i=\frac{\Omega_{(n+1)}}{16 \pi G}
\left(\frac{n}{2\kappa}\right)^n
 n \Omega_i\int_{\mathcal{B}_p}dV_{(p)}\;
 (1-V^2)^\frac{n-2}{2} R_i^2\,.
\eeq
The entropy is obtained by integrating the local horizon area as
\beq\label{Sbf} S=\frac{\Omega_{(n+1)}}{4G} \left( \frac{n}{2
\kappa} \right)^{n+1}\int_{\mathcal{B}_p}dV_{(p)}\; ( 1-
V^2)^{\frac{n}{2}} \,.
\eeq

\subsubsection*{Zero total tension}

Another useful quantity associated to the blackfold is the total
(integrated) tension, which in backgrounds with $R_0=1$ is defined as
\begin{equation}
\label{tension0}
\mbox{\boldmath$\TT$} = -
\int_{\mathcal{B}_p}dV_{(p)}\; (h^{\mu\nu}+\xi^\mu \xi^\nu)T_{\mu\nu}
\end{equation}
Using the stress tensor in \eqref{stress1},
\eqref{stress2}, we find that
\begin{equation}
\label{tension}
\mbox{\boldmath$\TT$}= \frac{\Omega_{(n+1)}}{16 \pi G}
\int_{\mathcal{B}_p}dV_{(p)}\; \left(\frac{n}{2\kappa}\right)^n
(1-V^2)^\frac{n-2}{2} \left(p-(n+p) V^2\right)\,.
\end{equation}
Using the explicit expressions \eqref{mass}, \eqref{angmom}, \eqref{Sbf}
and \eqref{tension} for the physical quantities of
the blackfold one finds that the relation
\begin{equation}
\label{sm1} (D-3) M = (D - 2) \left( \sum_i \Omega_i J_i + TS
\right) + \mbox{\boldmath$\TT$}
\end{equation}
generally holds for stationary blackfolds independently of whether the
extrinsic equations are satisfied or not. Indeed,
refs.~\cite{Harmark:2004ch,Kastor:2007wr} derived this expression for a
flat black $p$-brane of vacuum gravity in a $D$-dimensional space-time,
and following the methodology explained in \cite{Emparan:2009at}, one
extends it to an \textit{off-shell} blackfold identity by considering
$M$, $J_i$, $S$, $\boldsymbol{\TT}$ as functions of the worldvolume fields $X^\mu(\sigma)$.

On the other hand, for asymptotically flat black hole {\em solutions} of the
vacuum Einstein equations, the existence of one connected Killing
horizon generated by $\mathbf{k}$
implies the Smarr
formula
\begin{equation}
\label{sm2} \frac{D-3}{D-2} M =\sum_i \Omega_i J_i + TS\,.
\end{equation}
By comparing \eqref{sm1} and \eqref{sm2} we deduce
that asymptotically flat neutral blackfolds solutions satisfy a
zero-tension condition
\begin{equation}
\label{zeroten} \mbox{\boldmath$\TT$} = 0\,,
\end{equation}
$i.e.$ the total integrated tension vanishes when the blackfold equations
of motion are satisfied. In the
simplest cases where the blackfold equations \eqref{BFeqs} reduce to
only one equation, that equation is equivalent to \eqref{zeroten}.

\subsubsection*{Boundaries}

Black $p$-branes (and other fluid branes) may have `free' boundaries
without any boundary stresses. For  a neutral blackfold, vanishing
pressure at the boundary leads to \cite{Emparan:2009at}
\beq\label{bdryr0}
r_0\big\vert_{\partial\WW_{p+1}}= 0\,.
\eeq
Geometrically, this means that the horizon must approach zero size
at the boundary, so the horizon closes off at the edge of the
blackfold. In particular, for stationary blackfolds the condition
\eqref{bdryr0} means that $|\mathbf{k}|\to 0$ so the fluid velocity
becomes null at the boundary. This will happen when the fluid
approaches the speed of light at the boundary
\beq
V^2\big\vert_{\partial\WW_{p+1}}= 1\,.
\eeq
In Section \ref{ball}
examples of this will be presented. For stationary solutions the
extrinsic equations imposes a further condition namely that
$\perp^{\rho\mu}\partial_\mu r_0$ must vanish at the boundary at
least as quickly as $r_0$. This is verified as well in the
examples below.

\subsubsection*{Horizon geometry}

The blackfold construction puts, on any point in the spatial section
$\BB_p$ of $\WW_{p+1}$, a (small) transverse sphere $s^{n+1}$
with Schwarzschild radius $r_0(\sigma)$. Thus the blackfold represents a black hole with
a horizon geometry that is a product of $\BB_p$ and $s^{n+1}$ --- the product is
warped since the radius of the $s^{n+1}$ varies along $\BB_p$. The null
generators of the horizon are proportional to the velocity field $u$.

If $r_0$ is non-zero everywhere on $\BB_p$
then the $s^{n+1}$ are trivially fibered on $\BB_p$ and the horizon
topology is
\begin{equation}
\label{embedca}
(\mbox{topology of } \BB_p) \times s^{n+1}\,.
\end{equation}
However,  if $\BB_p$ has boundaries then $r_0$ will shrink to zero
size at them, resulting in a non-trivial fibration and different
topology. A simple but very relevant instance of this happens when
the topology of $\BB_p$ is that of a $p$-ball (see Sec.~\ref{ball}).
Then the horizon topology can easily be seen to be
$S^{p+n+1}=S^{D-2}$.

\subsubsection*{Stationarity after backreaction}

The formalism described above allows to find black hole solutions that
are stationary in a leading-order, test brane approximation. At the next
order, the gravitational backreaction of the object will modify the
solution. It is natural to ask if it will then remain stationary, or
instead acquire a slow time-dependence. We can expect that it will
indeed remain stationary as long as the leading-order solution does not
admit massless deformations (moduli) that do not break any symmetries.
Stationary solutions extremize an effective potential, \ie the
stationary action \eqref{BFact}. Backreaction effects will correct this
potential and lift the moduli, but a symmetry (\eg rotational symmetry)
can guarantee that the corrected potential still has an extremum near
the leading-order one. All of the solutions described in
Sections \ref{1folds}--\ref{nubc} fall in this class. However, if there are
massless deformations that do not break any symmetry, a non-trivial
potential will cause these moduli to roll. This is for instance the
situation in some multi-black hole solutions discussed in
Sec.~\ref{multibhs}.

\bigskip

Having reviewed the basic equations that govern the dynamics of
blackfolds, in particular in vacuum Einstein gravity, we are now in position to
explore specific solutions and their properties.

\section{Black one-folds}
\label{1folds}

The blackfold approach, when developed to leading order (probe
approximation) as we do in this paper, is almost trivial for zero-folds,
\ie\ small black holes that follow geodesic trajectories. The next
simplest case, black one-folds constructed out of black strings, is much
richer but still simple enough to derive results valid for generic
effective fluids and in backgrounds more general than we have assumed in
Sec.~\ref{equations}. These results are discussed in Sections \ref{sec:genres}
and \ref{sec:stat1folds1}.

In Sec.~\ref{sec:helical} we restrict ourselves to the context of Sec.~\ref{equations} and
discuss black one-folds in a Minkowski space background. It turns out to
be possible to find the general solution for a stationary black one-fold
in this background. This analysis also reveals a qualitatively new class
of black holes that confirm an outstanding open conjecture in
higher-dimensional gravity.

\subsection{General results}
\label{sec:genres}

We consider here the general analysis of one-branes valid for any background
and without imposing stationarity, the only assumption being that the effective
fluid is a perfect one, \ie dissipation effects are absent. A more extended analysis
including essentially equivalent results can be found in \cite{Carter:1989xk,Carter:1997pb}.
Later in Sec.~\ref{sec:stat1folds1} we restrict to stationary configurations.

Given a one-brane (a string) we can choose two orthonormal vectors tangent to its
worldsheet, $u$ and $v$, such that $u^2=-1$, $v^2=1$, $u^\mu v_\mu=0$. The first
fundamental form is
\beq
h_{\mu\nu}=-u_\mu u_\nu +v_\mu v_\nu
\eeq
and the integrability of the worldsheet submanifold requires that
$\pounds_u v$ remains tangent to the worldsheet, \ie
\beq
{\perp^\mu}_\rho[u,v]^\rho=0\,.
\eeq
If we take $u$ as the timelike unit vector that defines the velocity
field of the effective fluid, the stress tensor is
\beq
T_{\mu\nu}=\vep u_\mu u_\nu +P v_\mu v_\nu\,.
\eeq
Eqs.~\eqref{tsK} allow to immediately
write down the extrinsic equations \eqref{Carter}
as
\beq
\label{exgen}
{\perp^\rho}_\mu(\vep \nabla_u u^\mu +P\nabla_v v^\mu)=0\, .
\eeq

Considering now the intrinsic fluid equations, using the thermodynamic
relations $\TT ds=d\vep$, $sd\TT=dP$, $\vep+P=\TT s$ (where $\TT$ and
$s$ are the local temperature\footnote{Not
to be confused with the total tension $\mbox{\boldmath$\TT$}$ introduced
in \eqref{tension}.} and entropy density) one can easily rewrite the
continuity equation and the Euler equation as
\beq
{h_\mu}^\rho\nabla_\rho(su^\mu)=0\,,\qquad
{h_\mu}^\rho\nabla_\rho(\TT v^\mu)=0\,.
\eeq
The first one expresses that entropy is
conserved during the time evolution, as it must in a perfect fluid. The
second one refers to the spatial
uniformity of the temperature along the orbits of $v$.

\subsection{Stationary one-folds}
\label{sec:stat1folds1}

Now we restrict to stationary one-branes and describe two rather general
ways of deriving solutions. In a first approach we solve the Carter equation
\eqref{Carter} for generic one-branes in a class of backgrounds with $R_0=1$.
The second approach employs the action principle to obtain a very general
solution of stationary black one-folds that allows arbitrary $R_0$.

\subsubsection*{Solutions from the Carter equation}

The vectors $u$ and $v$ that we have introduced above define an
orthonormal local rest frame for the effective fluid. The
intrinsic fluid equations are solved for a stationary brane if the
velocity field $u$ is a timelike unit vector parallel to a Killing
vector of the form \eqref{kxichi}. We shall
assume that the timelike vector $\xi$ is unit-normalized on the
worldsheet, \ie $R_0=1$.
Let
$\zeta$ denote the unit-normalized spacelike vector orthogonal to $\xi$.
In the notation of \eqref{xitchiz},
\beq
\xi=\frac{\partial}{\partial t}\,,\qquad
\zeta=\frac{\partial}{\partial z}\,.
\eeq
Then we
can write
\beq
u=\cosh\alpha\xi+\sinh\alpha \zeta\,,\qquad
v=\sinh\alpha\xi+\cosh\alpha \zeta\,,
\eeq
where $\alpha$ is the boost relating the fluid local
rest frame to the frame of observers along orbits of $\xi$.
The boosted stress tensor is
\beqa
T_{\mu\nu}&=&(\vep\cosh^2\alpha+P\sinh^2\alpha)\,\xi_\mu\xi_\nu
+(\vep\sinh^2\alpha+P\cosh^2\alpha)\,\zeta_\mu\zeta_\nu\nonumber\\
&&+(\vep+P)\sinh\alpha\cosh\alpha \,\xi_{(\mu}\zeta_{\nu)}\,.
\eeqa
If $\xi$ is a unit-normalized Killing vector, then its orbits are
geodesics, $\nabla_\xi \xi=0$. If we assume further that $\zeta$ is
parallel-transported along $\xi$, $\nabla_\xi\zeta=0$, then
eqs.~\eqref{tsK} imply that
\beq\label{Kzz}
{K_{\mu\nu}}^\rho=\zeta_\mu \zeta_\nu \nabla_\zeta \zeta^\rho\,,
\eeq
and the extrinsic equations \eqref{Carter} are
\beq
(\vep \sinh^2\alpha +P\cosh^2\alpha)\nabla_\zeta \zeta=0\,.
\eeq
Thus, as long as the orbits of $\zeta$ are not themselves geodesics,
these equations require that
\beq\label{tanha}
\tanh^2\alpha=-\frac{P}{\vep}=c_T^2\,,
\eeq
and therefore the local fluid velocity $\tanh\alpha$ must be the same as the
propagation speed $c_T$ of elastic, transverse oscillations of the
string. Observe that this equation is the same as the zero tension
condition \eqref{zeroten}, since the pressure measured in the frame defined
by $(\xi,\zeta)$ is $\zeta_\mu\zeta_\nu T^{\mu\nu}$ and vanishes as a
consequence of \eqref{tanha}. In the present case, however, the
zero-tension condition does not require the solution to be
asymptotically flat (\eg it could be asymptotic to Kaluza-Klein, or
Kaluza-Klein monopole) nor indeed neutral. It only requires $R_0=1$.

If $\zeta$ on the worldsheet is proportional to a background
vector $\chi$ with orbit periodicity $2\pi$ and square norm $R^2$ on
the worldsheet, \ie\ $\zeta=\chi/R$, then this equation determines the
relation between $R$ and the angular velocity $\Omega$ relative to
orbits of $\xi$ as
\beq\label{rad1fold}
R\Omega=\sqrt{-\frac{P}{\vep}}\,.
\eeq

We have assumed $R_0=1$, which is a fairly restrictive condition on the
background. This can be easily relaxed in the following derivation of
the equations.

\subsubsection*{Solutions from the action}

We now use the action principle in order to perform a rather general
analysis of stationary black one-folds, with equation of state
\eqref{stress1}.

Given two background commuting Killing vectors $\xi$ and $\chi$ that are
tangent to the worldsheet, with norms on the worldsheet $R_0$ and $R$,
respectively, the entire worldsheet can be coordinatized by the
parameters along the orbits of these vectors. Therefore
the norms $R_0$ and $R$ are constant on the worldsheet. Under these
conditions, the action \eqref{BFact} (extended to include $R_0\neq 1$,
see \cite{Emparan:2009at}) can be trivially integrated over
the worldsheet and becomes\footnote{We neglect the inessential factor
$\beta$ of the time interval.}
\beq
I=2\pi R_0 R (R_0^2-\Omega^2 R^2)^{n/2}\,.
\eeq
Before proceeding to extremize this action with respect to variations of
$R_0$ and $R$, two notes of caution. First, the variational principle
reproduces the correct extrinsic equations only when the variations are
in directions transverse to the worldsheet. Thus if, for instance, we
consider a string wrapped on a uniform non-contractible circle direction
of radius $R$, no such variation can happen. Second, when $R_0$ is not
trivial, typically it can not be varied independently of $R$. Instead,
one may have to regard $R_0$ as a function of $R$. Taking this into
account, the equations that derive from this action
\beq
\frac{\partial I}{\partial R}+
\frac{dR_0}{dR}\frac{\partial I}{\partial R_0}=0
\eeq
are solved by
\beq
\Omega^2=\frac{R_0^2}{R^2}\frac{1+(n+1)
\frac{d\ln R_0}{d\ln R}}{n+1+\frac{d\ln R_0}{d\ln R}}\,.
\eeq
When $R_0=1$, this reproduces the result \eqref{rad1fold} applied
to the blackfold fluid \eqref{stress1}. More generally, this agrees with the
result in eq.~(5.5) of Ref.~\cite{Caldarelli:2008pz}, giving the balancing condition
for a class of metrics that includes the case of  black rings in AdS-space.

\subsection{All stationary one-folds in a Minkowski background: Helical
black strings and rings}
\label{sec:helical}

When the background is Minkowski spacetime we can find explicitly all
possible stationary one-folds. Stationarity with respect to asymptotic
observers requires us to choose the vector $\xi$ as the generator of Minkowski
time translations.
The orthogonal commuting vectors $\chi_i$ in \eqref{kxichi} must be
generators either of spatial translations or of rotations. Then this
situation falls under the assumptions of Sec.~\ref{sec:stat1folds1}
above. We can always choose the coordinate axes so that the curve that
the string lies along involves at most only one translational symmetry
$\partial_x$, which we shall assume has compact orbits of length $2\pi
R_x$. It is then convenient to introduce a coordinate $\phi_x=x/R_x$ with
periodicity $2\pi$. If the generators of rotations are
$\partial_{\phi_i}$ we can write the spatial
subspace of Minkowski spacetime in which the embedding of the string is
non-trivial as
\beq\label{subsp}
ds^2=R_x^2d\phi_x^2 +\sum_{i=1}^m \left(dr_i^2 +r_i^2 d\phi_i^2\right)
\eeq
and the string lies along the curve
\beq\label{curve}
\phi_x=n_x \sigma\,,\qquad r_i=R_i\,,\qquad \phi_i=n_i \sigma\,,\qquad
0\leq \sigma< 2\pi\,,
\eeq
with tangent unit vector
\beq
\zeta =\frac{1}{\sqrt{\sum_{a} n_a^2 R_a^2}}
\sum_{a} n_a \frac{\partial}{\partial \phi_a}\,.
\eeq
Here the vectors $\partial_{\phi_a}$  are evaluated on
the worldsheet, $r_i=R_i$. The indices run in
\beqa
i&=&1,\dots,
m\leq\left\lfloor\frac{D-1}{2}\right\rfloor=
\left\lfloor\frac{n+3}{2}\right\rfloor\,,\nonumber\\
a&=&x,1\dots,m\,.
\eeqa
The latter convention allows us to treat translations and rotations
jointly. Note, however, that the direction $x$ need not be always present.

The upper limit on $m$ is set by the rank of the spatial rotation group
in $D=n+4$ spacetime dimensions.
The $D-2m-2$ dimensions of space
that are not explicit in
\eqref{subsp} are
totally orthogonal to the string and we will ignore them.
They only play a role in providing, together with the $m$ directions
within \eqref{subsp} that are transverse to the curve, the $n+2$ dimensions
orthogonal to the worldsheet in which the horizon of the black string
is `thickened' into a
transverse $s^{n+1}$ of radius $r_0$.

The $n_a$ must be
integers in order that the curve closes in on itself. Without loss of
generality we assume $n_a \geq 0$. If we want to avoid multiple covering
of the curve we must have that the smallest of the $n_a$ (which need not
be unique), call it $n_{min}$, is coprime with all the $n_a$. Thus the
set of $n_a$ can be specified by $m$ positive rational numbers
$n_a/n_{min}$. Observe that $n_x$ is a winding number in the $x$ direction.

Obviously, if $n_x\neq 0$ and all
$n_i=0$ we recover a straight string.
If $n_x=0$ and all $n_i=1$ we obtain a circular planar ring along an orbit of
$\sum_i\partial_{\phi_i}$ with radius $\sqrt{\sum_i R_i^2}$. Since black rings where
exhaustively studied in \cite{Emparan:2007wm} we shall not consider this case, and
henceforth assume
either that $n_x\neq 0$ or that not all $n_i=1$.

Black 1-folds with $n_x\neq0$, and at least one $n_i\neq 0$, are
referred to as {\em helical black strings} (see Fig.~\ref{Helical}).  The black
objects that these blackfolds give rise to are, like black
strings, not globally asymptotically flat.

Black 1-folds with $n_x=0$ do give rise to asymptotically flat black
holes, and if at least two $n_i$ are non-zero and not both equal to one
then they are not planar rings. We refer to them as {\em helical black
rings} (see Fig.~\ref{Helical}).\footnote{They resemble the plasmid strings discussed in
\cite{BlancoPillado:2007iz}, but helical rings differ from the latter in that the profile of
the curve remains fixed and does not advance in time. Veronika Hubeny
has suggested the name `slinky' for these helical blackfolds, but
unfortunately this is a registered trademark, at least in three space
dimensions.}. Helical strings along an infinite, non-compact direction
can be obtained as limits of helical rings when one of the radii, say
$R_1$ becomes infinitely large. Formally, we have to make $n_1$ a
continuous variable, and send $R_1\to\infty$, $n_1\to 0$ while keeping
$R_x= n_1R_1$ finite, and introducing a new finite coordinate
$x=R_1\phi_1$.

\begin{figure}[t!]
\centering
\includegraphics[height=6cm]{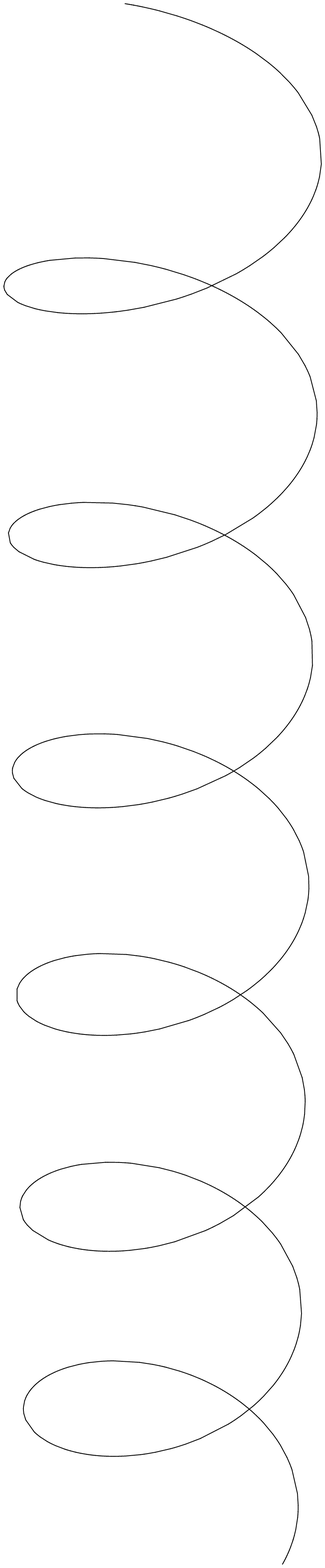}
\hspace{1.8cm}
\includegraphics[height=6cm]{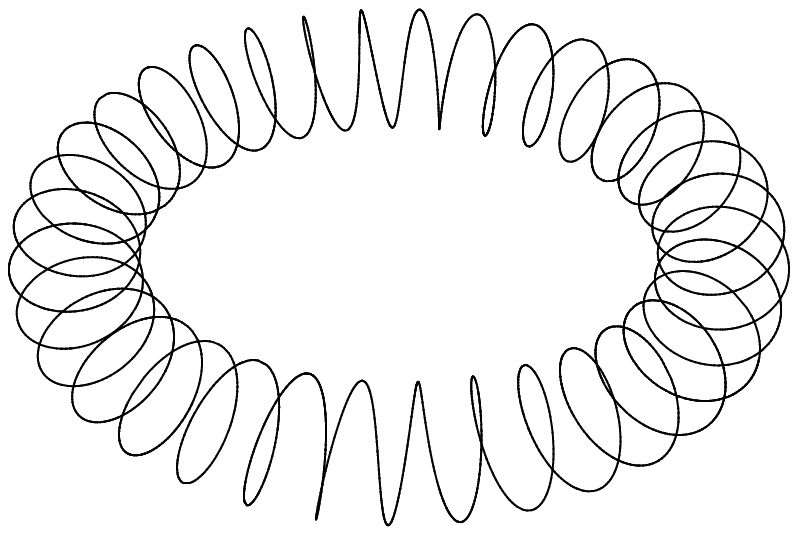}
\bf\caption{\it  Helical black strings (left) and helical black rings (right).}
\label{Helical}
\end{figure}

In this manner, all
possible stationary one-folds in a Minkowski background are classified into:
\begin{itemize}

\item Straight strings: $n_x\neq 0$, $n_i=0$.

\item Helical strings: $n_x\neq 0$, $n_i\neq 0$.

\item Planar rings: $n_x=0$, $n_i=1$.

\item Helical rings: $n_x=0$, and at least two $n_i>n_j> 0$.

\end{itemize}

\bigskip

The extrinsic curvature \eqref{Kzz} of the string is determined by
\beq
\nabla_\zeta \zeta=
-\frac{1}{\sum_a n_a^2R_a^2}\sum_i
n_i R_i\frac{\partial}{\partial r_i}
\eeq
which has constant norm along the curve and points towards the central axis
of the helix.

In general there will be a velocity $V$ along the string that can be
decomposed into components along the Killing directions giving
angular velocities $\Omega_i$ and a linear
velocity $V_x$ that we write as $V_x=R_x\Omega_x$. The Killing generator
of the worldsheet velocity field (and of the horizon) is
\beq
\mathbf{k}=\frac{\partial}{\partial t}+\sum_a \Omega_a
\frac{\partial}{\partial \phi_a}\,.
\eeq
The equilibrium condition \eqref{tanha} then fixes
\beq\label{vequil}
V^2=\sum_a \Omega_a^2 R_a^2=-\frac{P}{\vep}=\frac{1}{n+1}\,.
\eeq

We can write $\zeta$ in terms of the velocities
\beq
\zeta=\frac{1}{V}\sum_a|\Omega_a|
\frac{\partial}{\partial \phi_a}\,.
\eeq
Bear in mind that the ratios between angular velocities
must be rational
\beq\label{ratios}
\left|\frac{\Omega_a}{\Omega_b}\right|=\frac{n_a}{n_b}\quad  \forall a,b\,.
\eeq

We stress that, despite the presence of these angular velocities the
profile of the string remains fixed, so the configuration will not emit
any gravitational radiation. Indeed one expects that these solutions
will remain stationary also after backreaction effects are included: the
geometry along the curve is homogeneous, so gravitational
self-attraction will act homogeneously along all points of the curve
and, just like in the case of black rings, an increment in the velocity
will be enough to achieve equilibrium.

\subsubsection*{Maximal symmetry breaking and saturation of the rigidity theorem}

When a boosted black string is placed along the helix, the resulting
spacetime will have the isometry generated by
\beq\label{oneu1}
\sum_a n_a \frac{\partial}{\partial \phi_a}\,.
\eeq
However, the string breaks in general other $U(1)$ isometries of the
background, possibly leaving \eqref{oneu1} as the only spatial Killing
vector of the configuration.

To see this point, observe that any additional $U(1)$ symmetry must
leave the curve invariant, \ie the curve must lie at a fixed point of
the isometry. This is, the curve must be on a point in some plane in
$\mathbb{R}^{D-1}$, so rotations in this plane around the point leave it
invariant. Let us for simplicity consider helical rings (not strings)
and parametrize the most general possible helical curve in
$\mathbb{R}^{D-1}$ as a curve in $\mathbb{C}^m$, with
$m=\left\lfloor\frac{D-1}{2}\right\rfloor$, of the form
\beq
z_i=R_i e^{i n_i \sigma}\,,
\eeq
where possibly some of the
$n_i$ are zero.
In order to find a plane in which rotations leave the curve invariant we must
solve the equation
\beq
\label{aizi}
\sum_{i=1}^m a_i z_i=0
\eeq
with complex $a_i$, for all values of $\sigma$. This equation admits
a non-trivial solution only if some of the $n_i$ are equal to each other
(possibly zero).

Therefore, if (1) the string circles around in all of the
$m=\left\lfloor\frac{D-1}{2}\right\rfloor$ independent rotation planes,
\ie\ all the possible $n_i$ are non-zero, and (2) all the $n_i$'s are
different from each other, then the only spatial Killing vector of the
configuration is \eqref{oneu1}. In this case, we obtain an
asymptotically flat helical black ring with only one spatial $U(1)$
isometry.

The rigidity theorem of \cite{Hollands:2006rj, Moncrief:2008mr} requires the existence of
one such isometry, and we have found stationary black holes that have
exactly one spatial $U(1)$ in any $D\geq 5$. This provides constructive
proof of the conjecture made in \cite{Reall:2002bh} that posited
precisely this possibility.

\subsubsection*{Physical properties}

The total length of the string along the curve \eqref{curve} is
\beq
L=2\pi\sqrt{\sum n_a^2 R_a^2}\,.
\eeq
Using \eqref{vequil} and \eqref{ratios} we can write this as
\beq
L=\frac{2\pi}{\sqrt{n+1}}\frac{n_a}{|\Omega_a|}
\eeq
for any $a$.
The mass, angular momenta and entropy can be computed from \eqref{mass},
\eqref{angmom}, \eqref{Sbf} using
\eqref{vequil},
\beq
M=\frac{\Omega_{(n+1)}}{8G}(n+2)r_0^n\sqrt{\sum n_a^2 R_a^2}\,,
\eeq
\beq
J_a=\pm\frac{\Omega_{(n+1)}}{8G}\sqrt{n+1}\,r_0^n \,n_a R_a^2\,,
\eeq
\beq
S=\frac{\pi\Omega_{(n+1)}}{2G}\sqrt{\frac{n+1}{n}}\,r_0^{n+1}\sqrt{\sum n_a^2 R_a^2}\,.
\eeq
The sign of $J_a$ is the same as that of $\Omega_a$.
It should be noted that the $J_a$ are independent eigenvalues of the
angular momentum matrix only if condition (2) below eq.\ \eqref{aizi} is
satisfied for the non-vanishing $n_a$. In what follows we will assume this
is the case. When $n_x\neq 0$, $J_x/R_x$ is the linear
momentum along the translational direction of the helical string. We
have expressed these quantities in terms of the
thickness $r_0$, which itself is
\beq
r_0=\frac{n^{3/2}}{\sqrt{n+1}}\frac{1}{2\kappa}=\frac{n^{3/2}}{\sqrt{n+1}}\frac{1}{4\pi T}\,.
\eeq
These physical parameters satisfy simple identities,
\beq\label{id1}
J_a=\frac{n+1}{n+2}M\Omega_a R_a^2\,,
\eeq
\beq\label{id2}
M=(n+2)\sum_a J_a \Omega_a\,,
\eeq
\beq\label{id3}
\sqrt{\sum\frac{J_a^2}{R_a^2}}=\frac{\sqrt{n+1}}{n+2}M\,.
\eeq
The last one generalizes the expression first obtained in \cite{Emparan:2007wm}
that connects the radius of a ring to its mass and angular momentum. Using
\eqref{id1}, \eqref{id2} and the Smarr relation \eqref{sm2}, with
$D=4+n$, one can, for instance, derive
\beq
TS=\frac{n}{n+2}M\,.
\eeq

Neglecting numerical factors other than the $n_i$ the entropy behaves like
\beq\label{Sscaling}
S(M,J)\propto \left(\sum_a n_a |J_a|\right)^{-1/n}M^\frac{n+2}{n}\,.
\eeq
According to this formula the entropy decreases as we increase the $n_a$
while keeping M and $J_a$ fixed (although not arbitrarily since for very
large $n_a$ the strands of the string will be closer than $r_0$ and the
blackfold approximation will break down). This is reasonable; increasing
the $n_a$ makes the string longer, and thus thinner for fixed mass, which
reduces its entropy. Consequently, in order to obtain the maximal entropy
for a fixed number of non-zero angular momenta we should minimize the
$n_a$. For $p$ large angular momenta $J_1 \geq J_2 \geq \cdots \geq J_p$
this means that the $n=(1,2,\ldots,p)$ helical black ring has the maximal entropy.
For a single large angular momentum the planar black ring $n=(1,0,\ldots,0)$
maximizes the entropy.

\subsubsection*{Infinite rational non-uniqueness}

A helical black ring that
rotates along $m$ circles is completely specified by $r_0$, the radii $R_i$, and
the ratios $n_{i}/n_{min}$, which total $2m$ independent parameters. On the other
hand, the solution has $m+1$ conserved charges, $M$ and $J_i$. Therefore
for each helical black ring there is an infinite family of solutions,
parametrized by $m-1$ positive rational numbers, with the same mass and angular
momenta.

This infinite non-uniqueness is a new feature among vacuum
black holes with a single horizon.

\subsubsection*{No extremal helical black rings}

As discussed in \cite{Emparan:2009at}, one may consider black branes
where the small $s^{n+1}$ is rotating as the basis for blackfolds with
an internal spin. In order for these to remain stationary the internal
rotation directions must be isometries, so the blackfold must possess
additional $U(1)$s other than the one associated to the velocity field.
Then our results above imply that helical black rings cannot possess any
internal spin (\ie cannot be constructed out of Kerr or Myers-Perry
black strings).

It appears that the only way in which a neutral blackfold can be
extremal, \ie have a degenerate horizon, is if the rotation of the small
$s^{n+1}$ is itself at its maximum extremal value. Since MP black holes only
admit a regular extremal limit if all their spins are non-zero, it follows that
regularity of the
extremal limit of a blackfold requires that all the
$\left\lfloor\frac{n+2}{2}\right\rfloor$ internal spins are non-zero.
But a helical ring rotating in at least two planes breaks at least one
of the required isometries and therefore cannot satisfy the regular
extremality condition. It follows that extremal helical black rings are
not possible as regular asymptotically flat solutions of vacuum Einstein
gravity.

This impossibility of reaching a regular extremal limit applies to many
other blackfolds, since the bending of the brane does often break
isometries of the $s^{n+1}$. For instance, for all the toroidal
blackfolds $\T^p\times s^{n+1}$ with $p\geq 2$ discussed below, at least
one of the abelian isometries of the $s^{n+1}$ is broken by the toroidal
bending and therefore they do not admit a regular extremal limit.

Thus, while it cannot be discarded that new geometries for rotating
black strings and black branes exist that could overcome these
objections, it appears that studies of extremal near-horizon geometries
are severely limited for the description of ultraspinning vacuum black
holes.

\section{Solutions with odd-sphere horizon geometries}
\label{oddsphere}

A large family of solutions describes black holes in
$D$-dimensional flat space with horizon topology
\beq \label{oddSaa}
 \Big( \prod_{p_a=\rm odd} S^{p_a} \Big) \times s^{n+1}\,, \qquad \sum_{a=1}^\ell
p_{a}=p \,.
\eeq
In this case, the spatial section of the blackfold
worldvolume $\BB_p$ will be a product of odd spheres. As an
illustrative example, we will consider first the case of a single
odd-sphere $S^{2k+1}$. The simplest member of this family is the
thin black ring with horizon topology $S^1 \times s^{n+1}$, which was found and analyzed using blackfold methods in any spacetime dimension $D \geq 5$ in
\cite{Emparan:2007wm}.

\subsection{Black $S^{2k+1}$-folds} \label{singlesphere}

The starting point of the construction requires an embedding of the $S^{2k+1}$
sphere into a $(2k+2)$-dimensional flat subspace of $\R^{D-1}$, whose
metric is suitably
expressed as
\beq
\label{oddSab}
d r^2+r^2 \sum_{i=1}^{k+1}\left( d \mu_i^2+\mu_i^2 d \phi_i^2\right)
\,, \qquad \sum_{i=1}^{k+1} \mu_i^2=1
\eeq
where the $S^{2k+1}$ is parametrized by $k+1$ Cartan angles $\phi_i$
and $k$ independent director cosines $\mu_i$. In terms of these, the metric of a
$k$-dimensional sphere can be written as
\beq
\label{oddSac}
d \Omega^2_{k}=\sum_{i=1}^{k+1}d \mu_i^2
=\sum_{i,j=1}^k \left(\delta_{ij}+\frac{\mu_i\mu_j}{\mu^2_{k+1}}\right)d \mu_i d \mu_j
~.
\eeq
The additional $D-2k-3=n+1$ spatial dimensions of the background are
orthogonal to the
blackfold worldvolume and only play a spectator role in the analysis of
the blackfold equations. In the full black
hole solution, these directions, together with $r$, provide the
dimensions transverse to the worldvolume in which the
horizon is `thickened' into an $s^{n+1}$ of radius $r_0$.

We choose a gauge where the spatial worldvolume coordinates are
given by $\{ \mu_i | i=1,\ldots, k \}$ and the Cartan angles $\phi_i$,
$i=1,\ldots, k+1$. The embedding of the blackfold worldvolume
$\BB_{2k+1}$ is then described by a single scalar $r=R(\mu_1,\ldots,\mu_k)$.
We choose $r$ to be independent of the rotation angles since in order to
have a stationary blackfold we need the corresponding Killing vectors to
generate isometries of the worldvolume. The velocity field $u$ is
oriented along the Killing vector field
\beq \label{oddSad} {\bf k}=\frac{\d}{\d
t}+\sum_{i=1}^{k+1}\Omega_i \frac{\d }{\d \phi_i} ~.
\eeq
The
velocity function $V$ introduced in \eqref{kV} takes the form
\beq
\label{oddSae} V(\mu)^2=R(\mu)^2\sum_{i=1}^{k+1} \mu_i^2 \Omega_i^2
\,,
\eeq
where we have used that $R_i (\mu) =|\partial/\partial\phi_i|_{r=R}= R(\mu) \mu_i$.

These results amount to solving the intrinsic blackfold equations, \ie
obtaining the
velocity and pressure profiles of the effective fluid.
We can immediately read off the size $r_0(\mu)$ of
the small sphere $s^{n+1}$ using eq.~\eqref{r0}.

In order to obtain the extrinsic equations from the action \eqref{BFact}, in
addition to the velocity $V$ we need the
induced
worldvolume metric,
\begin{equation}
\label{oddSaf}
d s^2_{2k+1} = \sum_{i,j=1}^k \left[ \left( \delta_{ij}
+ \frac{\mu_i \mu_j}{\mu_{k+1}^2}\right) R(\mu)^2 + \partial_i R(\mu) \partial_j R(\mu)
\right] d \mu_i d \mu_j + R(\mu)^2 \sum_{i=1}^{k+1} \mu_i^2 d \phi_i^2
~.
\end{equation}

The generic expression for the action is complicated and will not be
presented here. An example of the general action and the
corresponding equations of motion can be found for the case of a
3-sphere in appendix \ref{s3}. In what follows, we proceed to
analyze a surprisingly simple but instructive case that allows an
explicit solution: a geometrically-round sphere
with constant radius $r=R$ that rotates with the same angular
velocity $\Omega$ in all $k+1$ directions $\phi_i$.

In this highly symmetric case, the angular rotation in Killing
vector,\footnote{Without loss of generality we have assumed that all
$\Omega_i$ are positive.}
\beq
\mathbf{k}=\frac{\d}{\d
t}+\Omega\sum_{i=1}^{k+1} \frac{\d }{\d \phi_i}\,,
\eeq
occurs along a diagonal $U(1)$ of the rotation group $SO(2k+2)$.
The velocity function $V$ in \eqref{oddSae}  simplifies to $R\Omega$ and the
stationary blackfold action \eqref{BFact} reduces to an $R$-dependent potential which reads
\beq \label{oddSag}
I[R]= \beta \Omega_{(p)} R^{p}\left( 1-
R^2\Omega^2 \right)^\frac{n}{2}\,, \qquad p=2k+1 \,.
\eeq
Varying this
potential with respect to $R$ we obtain the equilibrium condition
\beq
\label{oddSai} R=\sqrt{\frac{p}{n+p}} \frac{1}{\Omega}
\eeq
for a spherical blackfold. When $p=1$ it
agrees with the result in \cite{Emparan:2007wm} for black rings.

Notice that setting $R=\mathrm{const}$ directly in the action is a
consistent operation here: one obtains the same equations of motion if one 
puts $R$ a function of the worldvolume coordinates in the action, and then 
set $R$ to a constant in the derived equations. This is in fact easily verified,
since the extrinsic equations \eqref{Carter} for a round sphere reduce to the
equation
\beq \label{oddSaj} \frac{1}{R}\sum_{a=1}^{2k+1}
T^{aa}=0
\eeq
\ie the local brane tension vanishes, which immediately implies the
zero-tension condition \eqref{zeroten} on the integrated brane tension.
Physically this means that the tension of the static black $p$-brane is balanced
with the centrifugal force of rotation. Using $V= R\Omega$ in the expression
\eqref{tension}  for the total tension and requiring the integrand to be
zero yields \eqref{oddSai}.

This solution demonstrates the existence of thin regular
rotating black holes with horizon topology $ S^{2k+1} \times s^{n+1}$.
One may also consider
less symmetric solutions with the same horizon topology but non-equal
angular velocities. Then the radius $R$ becomes a non-trivial function
of the director cosines $\mu_i$. It obeys a second-order differential
equation, an example of which is discussed in appendix \ref{s3} for
$k=1$, that appears to require numerical analysis. Nevertheless, the
fact that the rotation group acts without fixed points and therefore the
$R_i$ do not vanish anywhere suggests that the centrifugal repulsion
should be able to balance the tension and thus non-round odd-sphere
solutions must be possible.

For round black $S^{2k+1}$-folds the physical properties can now be
computed straightforwardly from the general formulae of Section
\ref{equations}. Setting $p=2k+1$ and $V_{(p)} = R^p \Omega_{(p)}$
for the volume of a round $p$-sphere with radius $R$ we find
\begin{subequations}
\begin{equation}
\label{tbrthermoo2}
M=\frac{V_{(p)} \Omega_{(n+1)}}{16 \pi G} \, r_0^{n}(n+p+1)\,,
\end{equation}
\begin{equation}
\label{SToddS} S=\frac{V_{(p)} \Omega_{(n+1)}}{4G}  r_0^{n+1}
\sqrt{\frac{n+p}{n}} \spa T = \frac{n}{4\pi} \sqrt{ \frac{n}{n+p}}
\frac{1}{r_0}\,,
\end{equation}
\begin{equation}
\label{tbrthermo23}
J_i=\frac{1}{k+1} \frac{V_{(p)} \Omega_{(n+1)}}{16\pi G}\,R \,
r_0^{n}\sqrt{p(n+p)}\,,\qquad \Omega_i =  \sqrt{\frac{p}{n+p} }
\frac{1}{R} \spa i = 1 \ldots k+1 \, .
\end{equation}
\end{subequations}
For round-sphere solutions the thickness $r_0$ is a constant
independent of the worldvolume coordinates of the blackfold. The
expressions \eqref{tbrthermoo2}-\eqref{tbrthermo23}, satisfy correctly
the Smarr relation \eqref{sm2}. It is easy to see that these solutions
do not break any of the commuting isometries of the background.

\subsection{The general product of odd-spheres}
\label{prododd}

Returning to the more general case of odd-sphere products \eqref{oddSaa},
we can easily extend the previous discussion to obtain solutions of the
blackfold equations where $\BB_p$ is now a product of round
odd-spheres, each one labeled by an index $a=1,\ldots, \ell$.
Denoting the radius of the $S^{p_{a}}$ factor ($p_{{a}}=$odd) by
$R_{{a}}$ we take the angular momenta of the $a$-th sphere to be all
equal, \ie
\beq
\Omega_i^{({a})} = \Omega^{({a})}\,,\qquad \forall i=1, \ldots, \frac{p_{a}+1}{2}\,.
\eeq
We embed the product of $\ell$ odd-spheres
in a flat $(p+\ell)$-dimensional subspace of $\R^{D-1}$ with metric
\beq
\label{oddSba}
\sum_{a=1}^\ell \left( d r_{a}^2+r^2_{a} d \Omega^2_{(p_{a})}\right) \,,\qquad
\sum_{a=1}^\ell p_a = p
\eeq
and regard $r_a=R_a$ as the embedding scalars of our blackfold.
The worldvolume of the blackfold is a point in the transverse $\R^{n+2-\ell}$
subspace of $\R^{D-1}$. Hence, the dimension of the transverse space
is positive only when $\ell \leq n+2$. Given $n$, this inequality puts an
upper bound on the allowed number of spheres in the product.

By requiring $R_a$ to be constant functions on the worldvolume the stationary
blackfold action \eqref{BFact} reduces in this case to the potential
\begin{equation}
\label{oddSbb}
I[\{R\}] = \beta \prod_{b=1}^\ell  \Omega_{(p_b)} R_b^{p_b}
\left( 1  - \sum_{a=1}^\ell \left (R_a \Omega^{(a)}\right )^2 \right)^{n/2}
\end{equation}
which is the straightforward generalization of eq.\ \eqref{oddSag}.
Varying this potential with respect to each of the $R_a$'s we get a set of
$\ell$ equations, the solution of which gives the equilibrium conditions
\begin{equation}
\label{oddSbc}
R_a = \sqrt{ \frac{p_a}{n+ p} } \frac{1}{\Omega^{(a)}}
~.
\end{equation}

An especially simple case of this type of solutions is the
$p$-torus, where we set $p_{a} =1$, for all $a$. This choice gives
rise to blackfolds with horizon topology
\begin{equation}
\label{oddSbd}
\T^p \times s^{n+1}  \,,\qquad p \leq n+2
~.
\end{equation}
These solutions represent regular black holes rotating simultaneously along $p$
orthogonal $S^1$ cycles of the $p$-dimensional torus. Generalizing the black ring
analysis of \cite{Emparan:2007wm} we have constructed in appendix \ref{tori} explicitly 
the full perturbative metric of such black holes to leading order in $r_0/R_a$ using the 
method of matched asymptotic expansion and have verified that regularity of the perturbative 
solution requires the balancing condition \eqref{oddSbc}. It is noteworthy
that for  $n=1$ the hypergeometric functions appearing
in the first-order corrected near-horizon metric simplify
drastically, as observed already \cite{Emparan:2007wm} for the $n=1$ case in
the black ring family  $S^1 \times s^{n+1}$ and expected from the form of the
exact five-dimensional black ring solution. Here we find that the same simplification occurs for the
horizon topologies $\T^2 \times s^2$ in $D=6$ and $\T^3 \times s^2$ in $D=7$.

The leading order thermodynamics of a product of odd-spheres can be
computed again with the use of the formulae of Section
\ref{equations}. The expressions for $M, S$ and $T$ coincide with
the ones in \eqref{tbrthermoo2}, \eqref{SToddS}, provided we set
$V_{(p)} =\prod_{a} V_{(p_a)}$. The angular momenta and velocities
are given by the expressions
\begin{equation}
\label{tbrthermo23b}
J_i^{(a)}=\frac{2}{p_a+1} \frac{V_{(p)} \Omega_{(n+1)}}{16\pi G}\,R_a \, r_0^{n}\sqrt{p_a(n+p)}\,,
\qquad
\Omega_i^{(a)} =  \sqrt{ \frac{p_a}{n+p} } \frac{1}{R_a}
\end{equation}
where for each label $a$, the index $i$ runs from 1 to $(p_a+1)/2$.
Once again, the validity of the Smarr relation \eqref{sm2} can be
easily verified.

\section{Ultraspinning MP black holes as even-ball blackfolds}
\label{ball}

The blackfold equations do not admit solutions where $\BB_p$ is a
topological even-sphere in a Minkowski background. The tension at fixed
points of the rotation group cannot be counter-balanced by centrifugal
forces, so regular solutions of this type do not exist. Instead, as
announced in \cite{Emparan:2009cs}, there are solutions where $\BB_p$ is
an even-dimensional ellipsoidal ball $B_p$ with spatially varying
thickness $r_0$ that vanishes at the boundary of the ball. In this case,
$B_p$ has a `free' boundary without boundary stresses. Such
configurations are possible for black branes. The pressure, which is
proportional to $r_0$, goes to zero at the boundary and the horizon
closes off smoothly to produce a spherical horizon topology $S^{D-2}$.
We will see presently that these spherical-horizon solutions reproduce
{\it precisely} the physical properties of an ultra-spinning Myers-Perry
black hole with $p/2$ ultra-spins. This observation provides a highly
non-trivial check of the blackfold approach and shows that the method
remains sensible in situations with rotation fixed points ---in this
case, fixed points arise at the center of the ball.

To construct the solutions of interest we consider a planar $2k$-fold that spans the
$2k$-dimensional Euclidean subspace
\beq
\label{ballaa}
\sum_{i=1}^k \left( d r_i^2+r_i^2 d \phi_i^2 \right)
~ ,
\eeq
within $D=2k+n+3$ Minkowski spacetime.

Being flat, the $\mathcal{B}_{2k}$ worldvolume solves the extrinsic equations trivially.
In order to find a non-trivial solution of the intrinsic equations we set the blackfold
in rotation along the $\phi_i$ directions with corresponding constant angular velocities
$\Omega_i$. The worldvolume velocity field is then
\beq
\label{ballab}
V^i=\Omega_i r_i~~
\eeq
and the blackness conditions that solve the intrinsic equations give $r_0$ as a
function of the radial coordinates $r_i$
\beq
\label{ballac}
r_0=\frac{n}{2\kappa} \sqrt{1-\sum_{i=1}^k \left(\Omega_i r_i\right)^2 }
~ .
\eeq

Requiring $r_0$ to be real, or equivalently that the velocity field
does not exceed the speed of light, restricts the blackfold
worldvolume to the bounded region
 \beq
\label{ballad} \sum_{i=1}^k \Omega_i^2 r_i^2 \leq 1 ~,  \eeq which
defines an ellipsoidal even-ball. Approaching the boundary of the
ellipsoidal even-ball $r_0$ goes to zero, in accordance with
Eq.~\eqref{r0}, thus closing off the horizon smoothly. The velocity
field approaches there the speed of light.

A detailed calculation, that we defer to appendix \ref{app:ballsMP},
gives simple expressions for the physical properties of these solutions
\begin{subequations}
\beq \label{ballaea} M=\frac{r_+^n}{8 G}
\frac{(1+2k+n)\pi^{k+\frac{n}{2}}}{\Gamma(1+k+\frac{n}{2})}
\prod_{\ell=1}^k \Omega_\ell^{-2} ~, \eeq \beq \label{ballaeb}
J_i=\frac{r_+^n}{4G} \frac{\pi^{k+\frac{n}{2}}} {\Gamma\left(
1+k+\frac{n}{2}\right)} \Omega_i^{-1} \prod_{\ell=1}^k
\Omega_\ell^{-2} ~, \eeq \beq \label{ballaec} S = \frac{\pi}{2G}
r_+^{n+1} \frac{\pi^{k+\frac{n}{2}}}{\Gamma(1+k+\frac{n}{2})}
\prod_{\ell=1}^k \Omega_{\ell}^{-2} ~. \eeq
\end{subequations}
We have introduced a constant
\beq
\label{ballaf}
r_+=\frac{n}{2\kappa}\,,
\eeq
that corresponds to the thickness $r_0$ at the axis of rotation $r_i=0$.
Again we can verify that the total integrated tension
$\mbox{\boldmath$\TT$}$ vanishes. Note, however, that contrary to the
case of the odd-sphere blackfolds discussed before, here only the
integrated tension vanishes, not the local one.
As an illustration, Fig.~\ref{MPtension} shows the local brane pressure 
$\sum_{a=1}^{2k} T^{aa} $ ($i.e.$, minus the tension) as a function of
the radius for the case $k=1$ and $n=3$. The figure is
representative for the generic case, in that the pressure close to the
rotation axis is negative and positive for larger radii such that
the integrated value is zero.

\begin{figure}[t!]
\centering
\includegraphics[height=6cm]{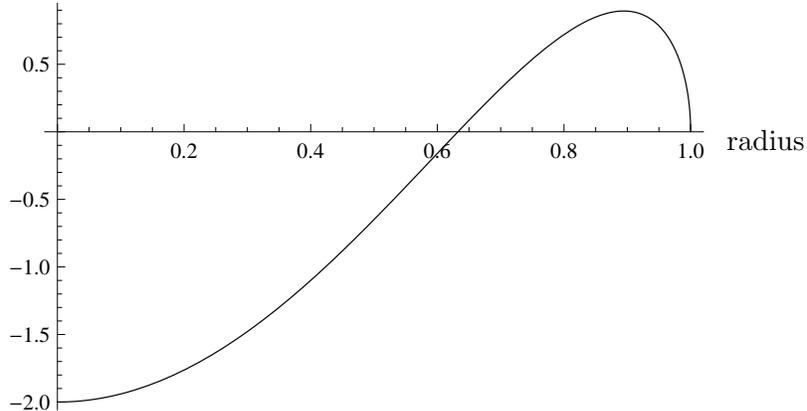}
\begin{picture}(0,0)(0,0)
\put(-6,100){radius}\put(-300,165){local brane pressure}
\end{picture}
\bf\caption{\it Local brane pressure ($i.e.$, minus tension) of the disc (2-ball) 
as a function of radius (in units of $1/\Omega$) for the case $n=3$. Integrating 
this brane pressure over the disc gives zero, as it should according to \eqref{zeroten}.}
\label{MPtension}
\end{figure}

Since the small sphere $s^{n+1}$ shrinks to zero size at the boundary,
the topology of the horizon is $S^{2k+n+1}=S^{D-2}$. This is the same as
the topology of the Myers-Perry black holes. Clearly, the blackfold
describes a highly distorted black hole whose size along the planes of
rotation is much larger than in the directions orthogonal to them. This
is also the shape of a Myers-Perry black hole solution in the limit of
$k=p/2$ large spins. In fact it was shown in \cite{Emparan:2003sy} that the
solution close to the center of rotation (in each of the $k$
rotation-planes) becomes approximately the metric of the flat static black
$p$-brane. This connects beautifully with our blackfold picture: We claim that
even-ball blackfolds describe precisely the ultraspinning regime of MP
black holes. The claim is quantitatively supported by the fact, shown in
appendix \ref{app:ballsMP},  that the physical magnitudes
\eqref{ballaea}-\eqref{ballaec} reproduce {\it exactly} those of
ultra-spinning MP black holes with $k=p/2$
ultra-spins in $D=n+p+3$ dimensions, once we identify the respective
values for the angular velocities and surface gravity. The
non-triviality of this exact match should be evident when one compares
the very different ways in which the quantities are obtained in each case.

The even-ball blackfold construction even manages to reproduce accurate
features of the shape of the horizon of the ultra-spinning MP black
hole. The thickness of the horizon transverse to the worldvolume of the
blackfold is
\beq
\label{ballai}
r_0= r_+\sqrt{1-\sum_{i=1}^k r_i^2\Omega_i^2}
~.
\eeq
For the MP black hole, the size of the transverse $S^{n+1}$ is
$r_+\cos\theta$, which is reproduced by making
\beq
\theta=\arcsin\left(\sqrt{\sum_i r_i^2 \Omega_i^2}\right)\,.
\eeq

\subsubsection*{No oddballs allowed}

We have seen that even-sphere blackfolds are not possible when the only
force available to balance the tension is the centrifugal rotation: the
blackfold collapses (tensionally, not gravitationally)
at the fixed points of the rotation. It is suggestive to think that the
even-ball is in fact the natural consequence of the flattening of the
even-sphere.

On the other hand, the converse of this reasoning leads to conclude that
oddball blackfolds are not possible, again if only rotation and tension
forces are present. The blackfold ball extends along rotation
\textit{planes} and therefore is naturally even-dimensional.

\section{Non-uniform black cylinders}
\label{nubc}

Another interesting example of stationary solutions with non-trivial embedding is provided
by inhomogeneous blackfold configurations that have cylindrical horizon topology
\beq
\label{cyliaa}
(\R\times S^1) \times s^{n+1}
~.
\eeq
In these configurations the cylindrical part of the blackfold worldvolume, $\BB_2$,
can be arranged to have varying $S^1$ radius and varying thickness $r_0$
resembling the Rayleigh-Plateau threshold mode of cylindrical streams of fluid
(see \eg \cite{Cardoso:2006ks,Caldarelli:2008mv}). These solutions
therefore provide an example of blackfolds with both non-trivial
worldvolume geometry \textit{and} non-uniform thickness $r_0$.

Compactifying one of the flat directions of the background and wrapping the non-compact
direction of $\BB_2$ around it gives stationary black hole solutions with compact horizon
topology $ ( \widetilde S^1\times S^1 )\times s^{n+1} $, where $\widetilde S^1$ is the compact
direction of the Kaluza-Klein (KK) background spacetime. The phase diagram of these solutions
exhibits some of the well known qualitative features of black strings in KK spaces
\cite{Harmark:2007md}. It will be discussed in the second half of this section.

An interesting class of generalizations, which will not be discussed here, involves
blackfolds with higher-dimensional worldvolumes $\BB_p$ where an
odd-sphere (or a product of odd-spheres) varies along one or more extra (compact or
non-compact) directions.

\subsection{Cylinders in non-compact flat space}
\label{flatspace}

Let us begin by considering $\BB_2$ as a surface embedded in a three-dimensional flat
subspace of $\R^{D-1}$ whose metric is conveniently expressed in cylindrical coordinates as
\beq
\label{cyliab}
d z^2+d \rho^2+\rho^2 d \theta^2
~.
\eeq
In this case $D=n+5$ and $\BB_2$ is parametrized by $z$ and $\theta$. Its embedding in
$\R^3$ is provided by the radial scalar $\rho=R(z)$ which can be a non-trivial function of $z$.
We allow the blackfold to rotate rigidly along $\theta$, so the velocity field $u$ is oriented
along the Killing vector field
\beq
\label{cyliad}
{\bf k}=\frac{\d}{\d t}+\Omega \frac{\d}{\d \theta}
~.
\eeq

With these specifications the stationary blackfold action \eqref{BFact} becomes
\beq
\label{cyliae}
I[R]=\int dz~ \LL[R(z),R'(z)]
=\int dz~ R(z) \left(1-\Omega^2 R(z)^2 \right)^{\frac{n}{2}} \sqrt{1+R'(z)^2}
~
\eeq
and the corresponding equation of motion is
\beq
\label{cyliaf}
RR''(1-\Omega^2 R^2)+(1+{R'}^2)\big( -1+(n+1) \Omega^2 R^2 \big)=0
~.
\eeq
Given a solution $R(z)$ of this equation the thickness of the blackfold $r_0$ is
\beq
\label{cyliafa}
r_0(z)=\frac{n}{2\kappa}\sqrt{1-\Omega^2 R(z)^2}
~.
\eeq

The uniform black cylinder is a simple solution of \eqref{cyliaf} where
$R$ is a constant function independent of $z$
\beq
\label{cyliag}
R(z)=R_c=\frac{1}{\sqrt{n+1}}\frac{1}{\Omega}
~.
\eeq
Since nothing depends on $z$, $R_c$ equals in this case the equilibrium radius of
a thin black ring (to verify this simply set $p=1$ in \eqref{oddSai}).

We can now examine if the differential equation \eqref{cyliaf} admits
more general solutions where $R$ is a non-trivial function of $z$. For
starters, we can look for such solutions perturbatively around the uniform
profile \eqref{cyliag} by setting
\beq
\label{cyliai}
R(z)=R_c+\epsilon \cos({\sf k}z)
~, ~~ \epsilon \ll 1
\eeq
as in the Rayleigh-Plateau analysis. Inserting this profile into \eqref{cyliaf} and expanding
up to first order in $\epsilon$ one finds that there is a solution with wavenumber ${\sf k}_c$ such
that
\beq
\label{cyliaj}
{\sf k}_c^2=\frac{2(n+1)^2}{n} \Omega^2
~~ {\rm or~equivalently} ~~
\left({\sf k}_c R_c \right)^2=\frac{2(n+1)}{n}
~.
\eeq
This is a threshold mode with period
\beq
\label{cyliak}
L=\frac{\pi\sqrt{2n}}{n+1}\frac{1}{\Omega}
~.
\eeq

The presence of the stationary inhomogeneous mode \eqref{cyliai} at
${\sf k}_c$ indicates that the cylindrical blackfold is unstable under mixed elastic and sound
wave perturbations with ${\sf k}<{\sf k}_c$. Such non-stationary perturbations can
be analyzed by considering the full set of intrinsic and extrinsic blackfold equations and
will be considered elsewhere.

Having established the presence of a stationary threshold mode we can now take a
step further to determine the inhomogeneous solutions of eq.\ \eqref{cyliaf} beyond
perturbation theory. For that purpose, it is convenient to reduce the order
of the differential equation by noting that the corresponding Hamiltonian density
\beq
\label{cylial}
\HH=\frac{\d \LL}{\d R'} R'-\LL=
-\frac{\left(1-\Omega^2 R(z)^2\right)^{\frac{n}{2}}R(z)}{\sqrt{1+R'(z)^2}}
\eeq
is a constant of motion, $i.e.$ a quantity independent of $z$. Accordingly,
we set
\beq
\label{cyliam}
\HH=-\frac{\nu_n-\nu}{\Omega}~, ~ ~ \nu_n=\frac{n^{n/2}}{(n+1)^{(n+1)/2}}
\eeq
where $\nu$ is a positive constant which we will call suggestively the non-uniformity
parameter. For $\nu=0$ we recover the uniform solution \eqref{cyliag}. We will discover
that increasing $\nu$ makes the cylindrical solution less and less uniform along $z$ until
we reach the critical value $\nu=\nu_n$ where $\HH=0$. At this point the solution
develops alternating tiny necks of vanishing size in $R$ and $r_0$.

\begin{figure}[t!]
\centering
\includegraphics[height=7cm]{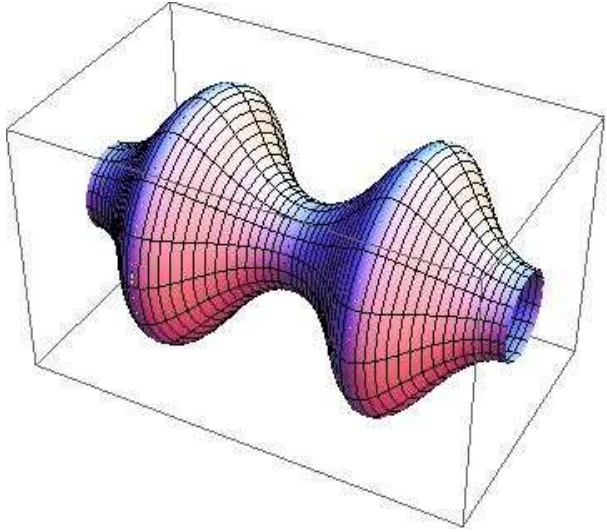}
\bf\caption{\it  A graphical depiction of the non-uniform black cylinder based on the
numerical evaluation of the blackfold equations for $n=1$ and $\nu=0.2$.}
\label{RPplot}
\end{figure}

We will solve the differential equation \eqref{cylial} numerically, but some of the main
properties of its solutions can be understood already by simple inspection. Rewriting
\eqref{cylial} as
\beq
\label{cylian}
\frac{d R}{d z}=\pm \frac{1}{\nu_n-\nu}
\sqrt{V_{\rm eff}(R)-(\nu_n-\nu)^2}
~, ~~ V_{\rm eff}(R)\equiv \Omega^2 R^2(1-\Omega^2 R^2)^n
\eeq
it becomes clear that there are non-uniform solutions with $R(z)$ oscillating between
turning points of the equation where $R'(z)=0$. At the turning points
\beq
\label{cyliao}
V_{\rm eff}(R)=(\nu_n-\nu)^2
~.
\eeq
For all $n\geq 1$, there are always two turning points, $R_-<R_+$, between which
$R$ oscillates with real first derivative provided the non-uniformity
parameter $\nu$ lies in the interval $[0,\nu_n]$. At $\nu=0$ the turning points
coalesce and one recovers the uniform solution \eqref{cyliag}. The oscillation
amplitude becomes maximum at $\nu=\nu_n$ where $R_-=0$ and
$R_+=\frac{1}{\Omega}$. In these solutions the thickness $r_0$ also
oscillates according to eq.\ \eqref{cyliafa}. It becomes maximum at $R_-$ and
minimum at $R_+$. One should bear in mind that the blackfold
approximation in principle breaks down near $R_-\approx 0$.

For each $\nu$ the period $L$ of the oscillation can be read easily from the differential equation
\eqref{cylian}. It is given by the integral expression
\beq
\label{cyliap}
L=2\frac{\nu_n-\nu}{\Omega} \int_{\Omega R_-}^{\Omega R_+}d x \,
\Big[ x^2 (1-x^2)^n-(\nu_n-\nu)^2\Big]^{-\frac{1}{2}}
~.
\eeq
The dimensionless quantities $\Omega R_\pm$ depend only on $n$ and $\nu$.

To illustrate these properties more clearly let us consider in more
detail the case $n=1$, \ie a cylinder black membrane in $D=6$. In this
case the turning points can be determined explicitly by solving a simple
quadratic equation
\beq
\label{cyliaq}
\Omega R_\pm =\sqrt{\frac{1}{2}\pm \sqrt{\nu(\nu-1)}}
~.
\eeq
We can solve the differential equation \eqref{cylian} numerically to find the profile
of $R(z)$ for any value of $\nu\in [0,\frac{1}{2})$. Fig.\ \ref{RPplot} is a three-dimensional
graphical depiction of the solution for $n=1$ and $\nu=0.2$.

\begin{figure}[t!]
\centering
\includegraphics[height=7cm]{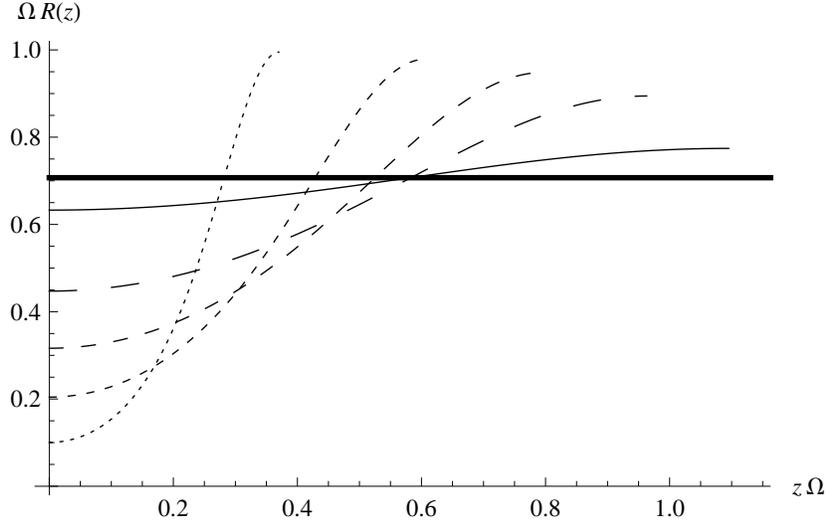}
\bf\caption{\it  Plot of the velocity $\Omega R(z)$
based on numerical solutions of eq.\ \eqref{cylian}
for $n=1$ and  fixed $\kappa$ and $\Omega$ .
Each curve represents $\Omega R(z)$ over a half-period. The curve
oscillates between a minimum $R_-$ and a maximum $R_+$. The thick solid
curve represents the uniform solution and has $\nu=0$. The remaining
curves, starting from the solid one to the shorter-dashed ones,
represent solutions with increasing values of $\nu$, respectively,
$\nu=0.01, 0.1, 0.2, 0.3, 0.4$.}
\label{n=1graph}
\end{figure}

Fig.\ \ref{n=1graph} depicts the result of the numerical evaluation over a half-period for six
values of $\nu$ for fixed $\kappa$ and $\Omega$.
As $\nu\to 0$ the profile of $R$ flattens and approaches that of the uniform
solution. Also observe that the numerical value of the half-period
($\frac{\Omega L_{num}}{2}\sim 1.1$) compares well at $\nu=0.01$
with the perturbative value \eqref{cyliak}
\beq
\label{cyliar}
\frac{\Omega L_{pert}}{2}=\frac{\pi}{2\sqrt 2}\sim 1.11072
~.
\eeq
As we increase $\nu$ the amplitude of $R$ increases and the period decreases.
Short necks are forming at two points: $(i)$ near the minimum of $R$, where
the diameter of the cylinder shortens, and $(ii)$ near the maximum of $R$, where
the thickness $r_0$ diminishes. As $\nu$ approaches its upper bound, $\nu_n$,
we enter a regime where the period $L(\nu)$ becomes comparable to $r_0$ and our
blackfold approximations break down. What happens in this highly non-linear regime
is an interesting problem.

\begin{figure}[t!]
\centering
\includegraphics[height=7cm]{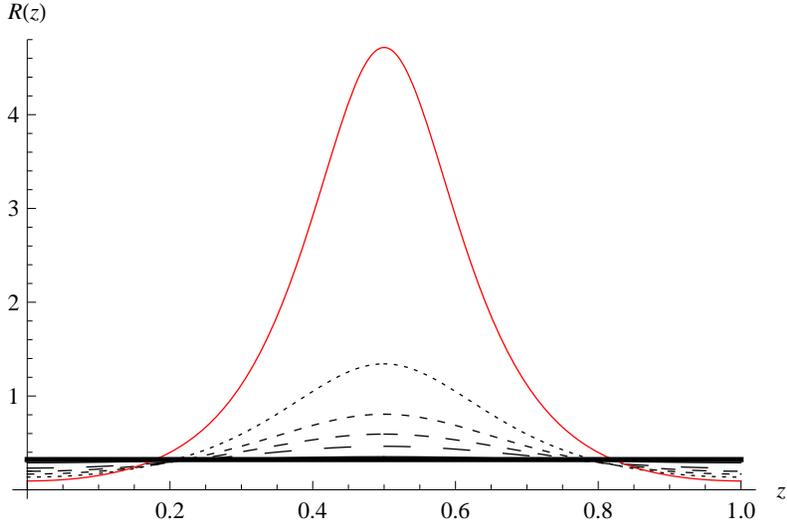}
\bf\caption{\it  Plot of the radius $R(z)$ as a function of $z$ for fixed $L$ and angular
momentum $J$, based on numerical solutions of eq.\ \eqref{cylian}, \eqref{cyliap},
\eqref{cyliat} where we chose $n=1$, $L=1$, $J=1$ and $G=1$.
The curve oscillates between a minimum $R_-$ and a maximum $R_+$. The thick solid
curve represents the uniform solution and has $\nu=0$. The remaining
curves, starting from the solid one to the shorter-dashed ones,
represent solutions with increasing values of $\nu$, respectively,
$\nu=0.01, 0.1, 0.2, 0.3, 0.4$. The red curve has $\nu = 0.48$
and serves to illustrate what happens near extreme non-uniformity assuming
the blackfold solution is trustworthy in that regime.}
\label{n=1graphFixedL}
\end{figure}

Another way of depicting these results is to plot
$R(z)$ versus $z$ for fixed $L$. This is shown in
Fig.~\ref{n=1graphFixedL}.
Naively, the infinite cylinder appears to reach a critical point
where it breaks off into an array of ultra-spinning discs. This breaking-off
reminds of the splitting of cylindrical streams of fluids into droplets under the
Rayleigh-Plateau instability.
A similar splitting of the horizon has been argued for
non-uniform black strings \cite{Kudoh:2004hs}. We will return to this issue in the next subsection.

The mass, angular momentum and horizon area of these solutions can be computed
using the general formulae. Defining these quantities per half-period we obtain the
integral expressions
\begin{subequations}
\beq
\label{cylias}
M=\frac{\pi \Omega_{(n+1)}}{4 G} \left( \frac{n}{2\kappa}\right)^n
\int_0^{\frac{L}{2}} (1-\Omega^2 R^2)^{\frac{n-2}{2}}(n+1-\Omega^2 R^2)R
\sqrt{1+{R'}^2} d z
~,
\eeq
\beq
\label{cyliat}
J=\frac{\pi \Omega_{(n+1)}}{4 G} \left( \frac{n}{2\kappa}\right)^n n \Omega
\int_0^{\frac{L}{2}} (1-\Omega^2 R^2)^{\frac{n-2}{2}}
R^3 \sqrt{1+{R'}^2} d z
~,
\eeq
\beq
\label{cyliau}
A=4GS=4\pi \Omega_{(n+1)}  \left( \frac{n}{2\kappa}\right)^{n+1}
\int_0^{\frac{L}{2}} (1-\Omega^2 R^2)^{\frac{n}{2}}
R \sqrt{1+{R'}^2} d z
\eeq
\end{subequations}
which can be evaluated numerically.

In order to compare different solutions, it is also instructive to
compute a set of dimensionless thermodynamic quantities defined as
\cite{Emparan:2007wm}
\beq
\label{cyliba}
j=c_j \frac{J}{M(GM)^{\frac{1}{D-3}}} \sim \frac{\ell_J}{\ell_M}~, ~~
a=c_a \frac{A}{(GM)^{\frac{D-2}{D-3}}}\sim \frac{A}{\ell_M^{D-2}}
\eeq
where $c_j$ and $c_a$ are conveniently chosen constants.
The $\kappa$ and $\Omega$ dependence of $M$, $J$ and $A$ follows easily by scaling.
Accordingly, for any $\nu\in (0,\frac{1}{2})$ one can show that the dimensionless area of the
cylindrical blackfolds behaves as
\beq
\label{cylibc}
a=f(\nu,n) j^{-\frac{2}{n}}
\eeq
where $f$ is a function of $\nu$ and $n$ that can be determined by solving the
differential equation \eqref{cylian} and evaluating the integrals \eqref{cylias}-\eqref{cyliau}.
We have performed this computation for several values of $n$. For $n=1$ the result has
been plotted in fig.\ \ref{ajplot}. As we increase $n$ the function $f$ develops a weak maximum
at intermediate values of $\nu$, but other than that it retains the main qualitative features
that are observed in fig.\ \ref{ajplot}.

\begin{figure}[t!]
\centering
\includegraphics[height=7cm]{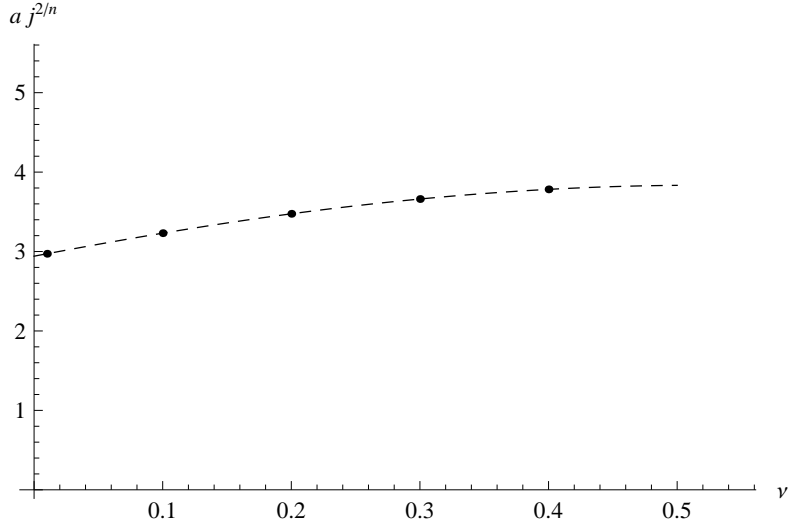}
\bf\caption{\it  A plot of the product $a j^{2/n}$ as a function of $\nu$ based on the numerical
solution of eq.\ \eqref{cylian} for $n=1$. $a$ and $j$ are respectively the dimensionless horizon
area and angular momentum. The dots represent the outcome of the numerical computation
with the choice $c_j=c_a=1$ in the definitions \eqref{cyliba} of $a$ and $j$.
The dashed line is an interpolation performed by {\sc mathematica}. Near the right end of this
plot $(\nu \sim 0.5)$ our approximations break down and our results should not be trusted.}
\label{ajplot}
\end{figure}

\subsection{Cylinders in Kaluza-Klein space}
\label{KKspace}

Compactifying one of the directions of the background and wrapping the cylindrical
blackfold around it we obtain a black hole with compact horizon topology
$ (\widetilde S^1\times S^1 )\times s^{n+1}$. In terms of the parametrization \eqref{cyliab}
the KK $\widetilde S^1$ is labeled by
\beq
\label{KKspaceaa}
z\sim z+2\pi \widetilde R
~.
\eeq
Everything we said about cylindrical blackfolds in the previous subsection continues
to hold in the compact case. The only difference comes from the constraints imposed by
the periodicity relation \eqref{KKspaceaa}. In particular, a non-uniform black cylinder
can now fit in the KK direction if the KK circumference is an integer multiple of the
period of the solution $L$, $i.e.$ if
\beq
\label{KKspaceab}
2\pi \widetilde R =LN~, ~ ~ N\in \N
~.
\eeq
Since $L$ is a function of $\nu, n$ \eqref{cyliap}, this implies a discretization of the allowed
values of $\nu$. In what follows, we will fix $N$ and $\widetilde R$ (equivalently we fix
$L$ from \eqref{KKspaceab}) and consider how the thermodynamics vary in terms of the
remaining parameters.

\begin{figure}[t!]
\centering
\includegraphics[height=7cm]{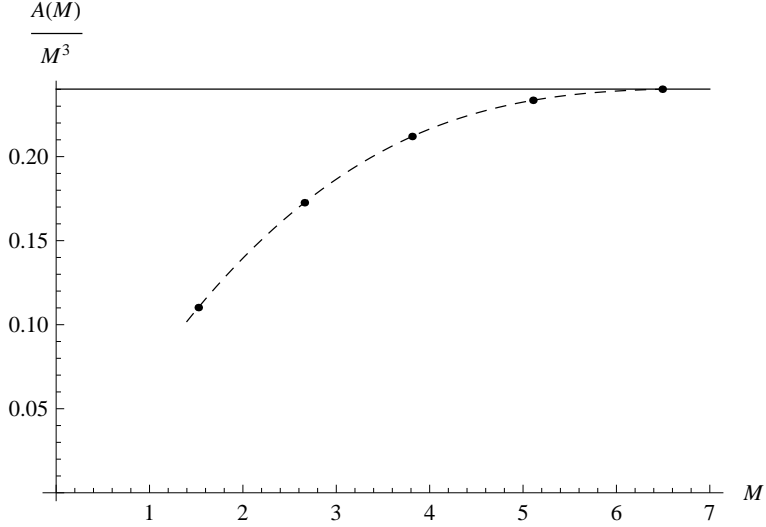}
\bf\caption{\it A plot of the rescaled horizon area $A/M^{\frac{n+2}{n}}$ as a function of the
mass $M$ for uniform and non-uniform black cylinders on a KK space. The specific plot
is based on a numerical evaluation of the stationary blackfold equations for $n=1$, $L=1$,
$J=1$ and $G=1$. The uniform phase is represented by the solid horizontal line. The
non-uniform phase is represented by the dashed line that emerges from the uniform branch
at the critical mass $M\sim 6.49$.}
\label{AMplot}
\end{figure}

With fixed $L$ the non-uniformity parameter $\nu$ becomes a function of $\Omega$.
This function $\nu(\Omega)$ can be determined by inverting the relation \eqref{cyliap}.
Then, $M$, $J$ and $A$ take the form
\beq
\label{KKspaceac}
M=\frac{f_M\left(\nu(\Omega),n\right)}{G \kappa^n \Omega^2}~, ~~
J=\frac{f_J\left(\nu(\Omega),n\right)}{G \kappa^n \Omega^3}~, ~~
A=\frac{f_A\left(\nu(\Omega),n\right)}{G \kappa^{n+1} \Omega^2}
\eeq
where $f_M$, $f_J$ and $f_A$ are functions of $\nu(\Omega), n$ that follow from the
expressions \eqref{cylias}-\eqref{cyliau}. Using the second equation in \eqref{KKspaceac}
to re-express $\kappa$ in terms of $J$ we can write $M$ and $A$ as
\beq
\label{KKspacead}
M=\frac{f_M\left(\nu(\Omega),n\right)}{f_J\left(\nu(\Omega),n\right)}\Omega J~, ~~
A=\frac{f_A\left(\nu(\Omega),n\right)}{\left[f_J\left(\nu(\Omega),n\right)\right]^{\frac{n+1}{n}}}
\left(GJ\right)^{\frac{n+1}{n}} \Omega^{\frac{n+3}{n}}
~.
\eeq
The $J$-dependence of these expressions is explicit. The $\Omega$-dependence
is more complicated and relies on the full solution of the stationary blackfold equations.
A numerical evaluation of this dependence has been performed for several values of $n$
and the result is represented by the dashed curve in fig.\ \ref{AMplot} for $n=1$, $L=1$ and
fixed $J=1$. For higher values of $n$, the curve continues to exhibit the same qualitative
features.

The uniform phase is represented by a straight solid curve in fig.\ \ref{AMplot}.
Re-evaluating the thermodynamics \eqref{cylias}-\eqref{cyliau} for the uniform cylinder
\eqref{cyliag} we find
\beq
\label{KKspaceae}
A=f_{A,u}(n) M^{\frac{n+2}{n}}J^{-\frac{1}{n}}L^{-\frac{1}{n}}
\eeq
where $f_{A,u}$ is the $n$-dependent coefficient
\beq
\label{KKspaceaf}
f_{A,u}(n)=\frac{8n}{n+2} \left( \frac{\Gamma\left(1+\frac{n}{2}\right)2^{n+2} (n+1)^{\frac{n+1}{2}}}
{n^{\frac{2n}{2}}\pi^{2+\frac{n}{2}}(n+2)^2}\right)^{\frac{1}{n}}
~.
\eeq

The non-uniform branch is emerging from this phase at a critical value of $M$ ($M\sim 6.49$
for $n=1$, $L=1$, $J=1$) and evolves towards smaller entropy and mass. The last data
point with $M\sim 1.52$ has $\nu=0.4$. As $\nu$ approaches its upper bound $\nu_1=0.5$
the gradients of the solution increase and our approximations break down. In the case of
$N=1$, a single blob of the non-uniform solution (see fig.\ \ref{n=1graphFixedL}) is wrapping the
$\tilde S^1$ direction $z$. Near the lower end of the non-uniform branch the solution
shows a clear trend towards pancaking around a point of the circle into a disc configuration
with vanishing $r_0$ at the edges. We analyzed such configurations in Section \ref{ball}
where we argued that they represent the ultraspinning regime of spherical horizon
Myers-Perry black holes. Hence, it seems appropriate to conjecture that the non-uniform
branch continues towards a topology-changing merger point where it meets the localized
branch of an ultraspinning Myers-Perry black hole in KK space%
\footnote{See Ref.~\cite{Kol:2007rx} for the leading order
correction to the thermodynamics of small Myers-Perry black in KK
space.}.
This picture is further
motivated by the corresponding phase diagram of black holes and strings in KK spaces
(see \cite{Harmark:2007md} for a review and references). In the present case, the blackfold
approach has enabled us to reduce the problem to a simple first order differential equation
like \eqref{cylian} that allows a straightforward analysis of a large part of the non-uniform
branch.

\section{Static minimal blackfolds}
\label{static}

If the blackfold is static then its velocity field is
\beq
u=\frac{1}{R_0}\xi\,,
\eeq
where $\xi$ is the timelike Killing vector of a static background.
The intrinsic equations are trivially satisfied with $r_0=n R_0/2\kappa$. Let
us assume that $R_0=1$. Then $u$ is a unit-normalized Killing vector
and therefore its orbits are geodesics. In this case eq.~\eqref{tsK} implies that
$u^\mu u^\nu {K_{\mu\nu}}^\rho=0$ and the equations \eqref{Carter} for a
perfect fluid become
\beq
Ph^{\mu\nu}{K_{\mu\nu}}^\rho=PK^\rho=0\,.
\eeq
Since the local fluid pressure $P$ is
always non-zero for a blackfold, the mean curvature vector $K^\rho$ must
vanish. Hypersurfaces that satisfy this condition are called minimal,
which is a bit of a misnomer since it is only guaranteed that they are
extremal. In any case we will refer to the corresponding black brane
configurations as minimal blackfolds.

Minimal surfaces are an intense field of study in mathematics and have a
wide range of applications. Here we discover that they can also be
relevant in the context of higher-dimensional black holes and branes.
Regularity implies that the worldvolume of static blackfolds spans a
sufficiently regular and non-intersecting minimal spatial submanifold.
Many interesting examples of such {\it embedded minimal surfaces} are
known and can be used as the basis for the construction of interesting
black hole configurations.
For instance, until fairly recently the only known minimal surfaces
embedded in three dimensions
were surfaces of revolution in
\beq
ds^2=dz^2+dr^2+r^2d\phi^2
\eeq
describing the plane,
\beq
z=\mathrm{const}\,,
\eeq
the helicoid
\beq
r=R\,,\qquad \phi=k z
\eeq
and the catenoid
\beq
r=R\cosh(z/R)\,.
\eeq
In 1982
Costa discovered a new minimal surface without self-intersections. This
is a non-compact surface with genus
one and three punctures. Since then more examples have been discovered.
They all provide new solutions for static non-compact black branes.

In Euclidean space there can be no compact embedded minimal surfaces.
Therefore it is not possible to construct a blackfold that gives a new
asymptotically flat static black hole. This is indeed consistent with
the theorem proved in \cite{Gibbons:2002bh} that the only such solution
in higher-dimensional vacuum gravity is the Schwarzschild-Tangherlini black
hole. All static neutral blackfolds in a Minkowski background are therefore
non-compact.

\section{Summary of horizon topologies and entropy ranking}
\label{phasescan}

In previous sections we presented a number of different solutions to the neutral
blackfold equations. Table \ref{horizontable}
summarizes the new types of asymptotically flat neutral black holes that
arise from the blackfold analysis of the previous sections. With a
possible M-theory motivation in mind we have included a list of horizon
topologies in $D\leq 11$ dimensions that we have so far found to be
allowed.

The list of Table \ref{horizontable} is only exhaustive of all possible
black hole topologies in $D=5$ \cite{Galloway:2005mf,Helfgott:2005jn}
(barring the possibility of a regular black lens
\cite{Hollands:2007aj,Evslin:2008gx,Chen:2008fa}), where in fact we have
obtained all possible blackfold solutions. But it provides already a
useful glimpse into the intricate structure of the phase space of
higher-dimensional black holes\footnote{See
\cite{Harmark:2009dh,Kleihaus:2009wh} for other recent work on possible
black hole geometries in higher-dimensions.}.

\begin{table}[t!]
\begin{center}
{\footnotesize
\begin{tabular}{|c|c|c|c|c|c|c|c|} \hline
$D=4$ & $D=5$ & $D=6$ & $D=7$ & $D=8$ & $D=9$ & $D=10$ & $D=11$ \\ \hline \hline
$S^2$  & $S^3$ & $S^4$ & $S^5$ & $S^6$ & $S^7$ & $S^8$ & $S^9$ \\ \hline \hline
 &  & ${B}_2 \otimes s^2$  & ${B}_2 \otimes s^3$ & ${B}_2 \otimes s^4$
 & ${B}_2 \otimes s^5$ & ${B}_2 \otimes s^6$ & ${B}_2 \otimes s^7$ \\
 &  &   &  & ${B}_4 \otimes s^2$ & ${B}_4 \otimes s^3$ & ${B}_4 \otimes s^4$ & ${B}_4 \otimes s^5$ \\
 &  &   &  &  &  & ${B}_6 \otimes s^2$ & ${B}_6 \otimes s^3$ \\
\hline \hline
& $S^1 \times s^2 $ & $S^1 \times s^3$ & $S^1 \times s^4$ & $S^1 \times s^5$
& $S^1 \times s^6$ &$S^1 \times s^7$ & $S^1 \times s^8$ \\ \hline
& & $ \T^2  \times s^2$ & $\T^2 \times s^3$ & $\T^2 \times s^4$ &
$\T^2 \times s^5$ & $\T^2 \times s^6$ & $\T^2 \times s^7$ \\ \hline
& &   & $S^3 \times s^2$ & $S^3 \times s^3$ & $S^3 \times s^4$
& $S^3 \times s^5$ &$S^3 \times s^6$ \\
& &   & $\T^3 \times s^2$ & $\T^3  \times s^3$ & $\T^3 \times s^4$
& $\T^3 \times s^5$ & $\T^3 \times s^6$ \\ \hline
& &   & & $S^1 \times S^3 \times s^2$  & $S^1 \times S^3 \times s^3 $
& $S^1 \times S^3 \times s^4$  & $S^1 \times S^3 \times s^5$  \\
& &   & &  & $\T^4 \times s^3 $ & $\T^4 \times s^4$ & $\T^4 \times s^5$ \\ \hline
& &  &  &  &      & $S^5 \times s^3$ & $S^5 \times s^4$ \\
& &  &  &  &     & $\T^2 \times S^3 \times s^3$ & $\T^2 \times S^3 \times s^4$ \\
& &  &  &  &      &  & $\T^5 \times s^4$ \\ \hline
& &  &  &  &     & $S^1 \times S^5 \times s^2$ & $S^1 \times S^5 \times s^3$ \\
& &  &  &  &     & $S^3 \times S^3 \times s^2$ & $S^3 \times S^3 \times s^3$ \\
& &  &  &  &     &  & $\T^3 \times S^3 \times s^3$ \\ \hline
& &  &  &  &     &  & $S^7 \times s^2$ \\
& &  &  &  &     &  & $\T^2 \times S^5 \times s^2$ \\
& &  &  &  &     &  & $S^1 \times S^3 \times S^3 \times s^2$ \\ \hline
\end{tabular}}
\end{center}
\bf\caption{\it A partial list of possible horizon topologies for
asymptotically flat neutral black holes in $D=4$ to $D=11$ dimensions
based on the currently available solutions of the corresponding
blackfold equations. $B_{2k}$ in the second row denotes the
$2k$-dimensional filled-ellipsoids of Section \ref{ball} that
appear in the blackfold description of ultraspinning MP black holes.
$\otimes$ denotes a warped product geometry so the topologies in the
second row are actually the same as in the first row. Capital (small)
letters
denote the part of the horizon with comparatively larger (smaller)
characteristic scales.}
\label{horizontable}
\end{table}

The first row includes the exactly known Kerr black hole in four
dimensions (the only phase allowed by uniqueness theorems), and
its higher-dimensional generalizations, the MP black holes. The second
row, which begins from $D=6$ and goes up, includes the possible
ultraspinning limits of the MP black holes as even-ball blackfolds of
the corresponding dimensionality. The topology of the horizons in the
first and second row are therefore the same.
The third row describes black ring
solutions in any dimension $D\geq 5$. The five-dimensional black ring
solution is known exactly \cite{Emparan:2001wn}. In higher dimensions
black ring solutions have been constructed perturbatively in the
ultraspinning limit \cite{Emparan:2007wm}. In this row one can also
package the new less symmetric helical black rings that we presented in
Section \ref{1folds}. The fourth row comprises of black two-tori. A
perturbative solution for the metric of this type of black holes appears
in appendix \ref{tori}. In subsequent rows increasingly complicated
black holes appear with horizon geometries that include both tori and
odd spheres. In previous sections we focused on round odd-sphere
solutions with equal angular momenta, but we pointed out that solutions
with unequal angular momenta are also possible (for a preliminary
discussion see appendix \ref{s3}). They should also be included in this
list. Not surprisingly, as we increase the spacetime dimension an
increasingly larger set of possibilities opens up.

The physical properties of the new solutions can be computed in the
ultraspinning limit using the blackfold methodology. It is interesting
to compare the properties of different solutions, $e.g.$ their
entropies. Under the assumption that the sizes of the compact directions
in ${\cal{B}}_p$ are all of the same order $R$, it was shown in
great generality in \cite{Emparan:2009at} that, for
fixed mass, the
rescaled entropy $s$ scales with the dimensionless angular momentum $j$ as
\begin{equation}
\label{sj}
s(j) \sim j^{-\frac{p}{D-p-3}}
~.
\end{equation}
This scaling relation shows that for a given number of ultra-spins, the
blackfold with smallest $p$ is the one that is entropically favored.
Physically, objects with smaller values of $p$ are thicker at given
mass and hence cooler. Since the entropy at constant mass is inversely
proportional to the temperature, it follows that thicker objects, with lower
$p$, have higher entropy. It must be recalled that an MP black hole with
$k$ ultraspins is not to be regarded as a $0$-fold, but rather as a
$2k$-fold.

Accordingly, the most entropic solutions among blackfolds with the same
number of ultra-spins are black rings, which will be helical if more
than one ultra-spin is involved. Since the helical ring with the shortest
length is entropically favored, the helical ring with maximal entropy is the
$n=(1,2,...,p)$ configuration, as discussed in Section \ref{sec:helical}.

\section{Multiple black holes from blackfolds}
\label{multibhs}

The blackfold approach allows to easily construct multi-black hole
configurations. Since the gravitational interaction of the blackfold is
neglected to leading order, one simply superposes several blackfolds in
a given configuration. As long as the separation between them is larger
than $r_0$ (the largest of the thicknesses involved), a
solution is obtained if all the blackfolds involved solve separately the
blackfold equations.

These solutions often possess moduli and therefore are
particularly sensitive to corrections from gravitational backreaction.
For instance, it is obvious that two parallel flat branes solve the
leading order, test-brane approximation, but they will attract and move
towards each other as soon as the gravitational interaction is turned
on. This is because the distance between the branes is a modulus and
changing the distance does not break any symmetry. In contrast in a
black Saturn with a central black hole and a surrounding black ring, the
configuration does not admit any deformation of the relative position of
the two objects that does not break any of its symmetries. Thus, when
gravitational interaction is turned on, the gravitational energy will be
extremized in the symmetric configuration and the configuration will
remain in equilibrium (although presumably an unstable one).

Black Saturns \cite{Elvang:2007rd} are the simplest configurations that
can be obtained in this way, and indeed were analyzed in this manner in
\cite{Elvang:2007hg}. They admit generalizations involving the different
new solutions we have found. For instance, a circular black ring may be
surrounded (but not linked) by a helical black ring. The entire
configuration also admits a central black hole\footnote{We thank
Veronika Hubeny for suggesting these generalized helical black
Saturns.}. Observe that the black holes involved in these
configurations can only rotate in the Killing directions that
are not broken by the helical ring.
In dimensions $D\geq 7$ an odd-sphere blackfold can be in
stationary equilibrium with a central black hole, or with other
concentrical blackfolds%
\footnote{See for example Refs.~\cite{Iguchi:2007is,Evslin:2007fv}
for di-rings}. Multiple blackfolds, each one lying in an
independent submanifold of the background, which generalize the bicycle
black rings of \cite{Elvang:2007hs,Izumi:2007qx}, are also possible.

In this discussion the values of the temperature and angular velocities
can be different among the several blackfolds involved. At the classical
level, there is no problem in having, say, a central black hole and a
surrounding black ring with different temperatures and angular
velocities. However, in some instances, \eg close to a merger between
phases, the equality of these quantities among different objects is
relevant. In ref.~\cite{Emparan:2007wm} it was suggested that `pancaked
black Saturns' in $D\geq 6$ consisting of a central ultraspinning black
hole and a surrounding black ring might exist with equal angular
velocities and temperatures, and could provide a way of connecting
different phases.

We now argue that this is not the case. Focusing on the case with a single
spin, an ultraspinning black hole corresponds to a blackfold disk, and
the black ring to a circular string. If the rotation velocity is for
both of them $\Omega$, then the radii of the two objects are related as
\beq
R_\mathrm{ring}=\frac{1}{\sqrt{D-4}\,\Omega}<R_\mathrm{disk}=\frac{1}{\Omega}\,,
\eeq
(cf.~\eqref{oddSai} with $p=1$, $n=D-4$ for
the ring, and \eqref{ballad} with $k=1$ for the disk). The inequality is
strict since $n\geq 2$, so the circular ring would have to be inside
the disk. Therefore the configuration is impossible, and ultraspinning
black holes cannot be in full thermodynamical equilibrium with a
surrounding black ring. As a consequence, the dashed phases and connections in
figure~6 of ref.~\cite{Emparan:2007wm} cannot be realized, at least when the
spin is sufficiently large.

\section{Discussion and open problems}
\label{conclusions}

In this paper we have studied new stationary configurations of neutral
black holes in dimension $D\geq 5$ by regarding them as thin black
branes that wrap a submanifold of a background spacetime. The effective
worldvolume theory that describes the long-wavelength dynamics of the
black brane was formulated very generally in \cite{Emparan:2009at}, and is quite
similar to the effective Dirac-Born-Infeld theory for D-branes in open
string theory, or the Nambu-Goto effective description for
Nielsen-Olesen vortices. In our opinion, the blackfold approach should
be regarded and judged in much the same way as one does in the case of
these other effective theories, both in terms of its validity and of its
utility. Blackfolds provide the leading order description of objects for
which an exact account is very probably out of practical reach.
Corrections to this leading order are more often than not very
complicated too, but unless there is good reason to do so, one does not
doubt the validity of this approximation to a full, physical solution
describing the object in the complete theory.

There is however one significant respect in which blackfolds differ from
DBI branes or NG strings (or indeed any other dynamical branes that we
are aware of): the worldvolume theory of blackfolds features proper
hydrodynamical behavior, in the sense of requiring local thermodynamical
equilibrium. In contrast, NG strings have constant energy density and
pressure, so their intrinsic dynamics is trivial, while the
worldvolume dynamics of DBI branes is a nonlinear electrodynamics that
does not involve thermal features. This is the main reason that in the
blackfold theory configurations in stationary equilibrium are
particularly singled out.

We hope the results of this paper demonstrate that this is indeed a
powerful tool for the identification of new solutions and their
properties. But its utility should not be reduced to only describing
novel classes of stationary black holes, but also to analyzing their
dynamics in specific physical situations. We now discuss these issues as
well as some interesting directions for future research that this
approach opens up.

\subsubsection*{\it The metric at all length scales}

The blackfold approach as pursued in this paper might be regarded with
scepticism since, although it is claimed that new black
holes are found, no explicit black hole metric appears to be produced.
Expressed in this crude form, this criticism is unwarranted.
First, let us emphasize again the similarity to the fact that in
general a solution of the DBI action does not provide
an explicit solution to the full open string theory (indeed quite often
it is not even known how to solve string theory in the backgrounds where
this effective theory is applied), and a similar situation occurs
for vortex strings and their Nambu-Goto description. Second, it is not
quite true that no metric for the new black hole is given. It actually
is, to leading order: far from the black hole, it is the background
metric, with a submanifold singled out as the location of the blackfold;
and near the black hole, it is the metric of a boosted black brane.

Nevertheless we admit that there is a point in this criticism, since
traditionally black holes have been regarded as embodying a non-trivial
geometry, and it might be desirable to see how a new metric is obtained
for the new black hole, at least in principle. Indeed the first
application of the blackfold methodology included a long and detailed
analysis of the next-to-leading order metric for higher-dimensional
black rings \cite{Emparan:2007wm}. In appendix \ref{tori} we perform
this analysis for black tori of general dimension.

As explained in detail in \cite{Emparan:2007wm} (following
\cite{Harmark:2003yz,Gorbonos:2004uc}) and \cite{Emparan:2009at}, the
method of matched asymptotic expansions (MAE's) systematically produces
an explicit solution for the geometry of the black hole spacetime at all
scales, including the region near the horizon, with the effects of the
bending of the black brane in an expansion in $r_0/R$. It should be
appreciated that the leading order MAE is a rather involved technical
task even in the very symmetric situations that we have studied, a fact
that emphasizes once again the virtue of having a universal,
long-distance effective theory that captures in a simple manner most of
the physically interesting features of the solution.

Nevertheless, it would be interesting, and possibly quite informative,
to perform a MAE in less symmetric examples where $r_0$ is a non-trivial
function of the worldvolume coordinates. The even-ball blackfolds that
describe the ultraspinning regime of MP black holes is such an example.

It would also be very interesting to go beyond the leading order in
effective field theory and MAE's. In the effective field theory side
this would require a systematic derivation of the blackfold equations
which would go beyond the leading order approximations to account for
the effects of dissipation, internal structure and gravitational
self-force. From the point of view of MAE's one would have to setup a
general expansion scheme as in \cite{Gorbonos:2004uc}.

\subsubsection*{\it The phase diagram of higher-dimensional black holes}

We have uncovered large new classes of higher-dimensional black hole
solutions in a specific ultraspinning regime (see table
\ref{horizontable} for a partial list in $4\leq D\leq 11$ dimensions).
The new solutions are part of a multi-dimensional `phase diagram' where
we can plot solutions at a fixed spacetime dimension with the axes being
their asymptotic charges and entropy. In this diagram there are
generically regions where several phases co-exist with the same
asymptotic charges providing new examples of black hole non-uniqueness
in higher-dimensional gravity. Indeed we have shown that helical black
rings give rise to an infinite non-uniqueness, labelled by rational
parameters, in all ultraspinning regions of the phase diagram. In these
regions, helical black rings have the highest entropy among phases with
connected horizons.

It would be very interesting to see how the new phases place themselves
into the general phase diagram and to trace them away from the blackfold
regime towards the non-linear regime of dynamics where different phases
typically merge. Obtaining analytic information about the non-linear
regime is a daunting task, but it may be possible to obtain valuable
information by targeting specific examples numerically.

A more mundane task would be to aim for a qualitative global picture of
the phase diagram extrapolating away from the available phases on the
basis of well educated assumptions. Ref.~\cite{Emparan:2007wm} provides
an example of such an approach. In that paper the discussion was limited
to higher-dimensional black holes with a single spin. It would be
desirable to generalize it to the case of black holes with more angular
momenta and obtain a similar picture that involves the new solutions
presented in this paper.

Another example of such extrapolations was briefly discussed in section
\ref{KKspace} in the context of black cylinders in KK space. In
fig.~\ref{AMplot} we can track the behavior of the non-uniform
ultraspinning black cylinder branch by solving the blackfold equations
up to a point where a thin neck appears in the blackfold worldvolume
geometry and our approximations break down. In the same phase diagram
one expects an additional curve that describes an ultraspinning MP black
hole localized in the KK direction. It would be interesting to determine
the thermodynamics of this phase for small mass in perturbation theory.
A natural expectation, which is corroborated by the behavior of the
non-uniform black cylinder solutions in Section \ref{KKspace}, is that
there is a topology-changing merger point where the non-uniform black
cylinder branch meets the localized MP branch.

\subsubsection*{\it Dynamical aspects: Stability and time-dependence}

Many of the emerging new solutions of higher-dimensional gravity exhibit
regions of instability in the blackfold regime. For example,
ultraspinning MP black holes and thin black rings have been argued
\cite{Emparan:2003sy} to be unstable under GL-type instabilities. Corresponding
statements can be made for more general blackfolds. As a matter of fact,
the blackfold approach does easily capture this instability for a
generic blackfold in a regime in which the wavelength $\lambda$ of the
instability lies in the range $r_0\ll \lambda \ll R$. The instability
appears in the intrinsic sector of the blackfold equations as an
unstable sound mode of the effective worldvolume fluid
\cite{Emparan:2009at}.

This is one example where one can decouple the extrinsic equations from
the stability analysis of the intrinsic sector. Conversely, there are
situations where the extrinsic stability can be analyzed while
guaranteeing that the intrinsic equations remain solved. A simple
instance is the study of stability against variations of the radius of
the round odd-sphere blackfolds of Sec.~\ref{oddsphere}. These solutions
extremize a potential $V=-I/\beta$ where $I$ is the action \eqref{oddSag} for
the stationary configurations. This $V$ is \textit{minimized} by these
solutions, implying that they are stable to variations of $R$ (this was
in fact already known for black rings \cite{Elvang:2006dd}). An explicit
analysis of time-dependent perturbations confirms this result. This, and
a more general investigation of instabilities and time-dependent physics
for the new blackfold solutions presented in this paper will be
discussed elsewhere.

\subsubsection*{\it Generating more solutions}

The construction of higher-dimensional blackfolds in this paper was
based on the black brane solution. It is possible to generalize this
construction in several interesting ways. Internal ultra-spins along the
transverse $(n+1)$-sphere, with rotation parameter $a\propto J/M\gg r_0$
can be incorporated by using the MP black branes as the starting point
of the construction. Small internal spins, with $a\lsim r_0$, can also
be incorporated easily to leading order since their effects only enter
at a higher order in the expansion in $r_0/R$. An element of novelty is
introduced by the zero-modes that appear at discrete values of the
rotation \cite{Dias:2009iu} which, when present, should be included as part of
the set of collective coordinates of the blackfold.

In general, the formalism can be applied easily to any brane whose
effective stress tensor is known. One example are blackfolds based on
the `lumpy' black branes of \cite{Gubser:2001ac,Wiseman:2002zc}, which
are black brane solutions branching off the regular black brane branch
at the threshold of the GL mode. The resulting solutions, coined `lumpy
blackfolds' in \cite{Emparan:2009at}, have horizons that are
inhomogeneous on the small scale $r_0$.

Another set of interesting examples arises with blackfolds based on
black branes constructed out of the exactly known black ring solution in
five dimensions by adding extra flat dimensions to the metric. Such a
ringy black string is in fact the exact solution that corresponds to the
six-dimensional uniform black cylinder of Sec.~\ref{nubc}.
Bending it on a transverse $S^1$ in six dimensions will produce a
blackfold with horizon topology $S^2\times S^1_{ring}\times
S^1_{extra}$. With generic spin along $S^1_{ring}$ the new solution can
now probe a different regime of the ultraspinning $\T^2$-blackfolds of
Section \ref{prododd}. Several generalizations of this construction can
be envisioned.

As explained in \cite{Emparan:2009at}, essentially the same formalism as
in this paper can be employed to study neutral blackfolds in any
background with characteristic length scale $\gg r_0$, independently of
what theory the background is a solution to as long as it contains the
Einstein-Hilbert sector. With charged blackfolds the situation is
different and will be discussed in detail elsewhere. One can use the
approach, for example, to uncover new single- or multi-charged black
hole solutions in supergravity theories that arise in 10D string theory
or 11D M theory and their compactifications. These will be discussed in a future
work.

\section*{Acknowledgments}
RE, TH and NO are grateful to the Benasque Center for Science for
hospitality and a stimulating environment during the Gravity Workshop in
July 2009, and they thank the participants there for very useful
feedback and discussions. We also thank Tobias Colding for useful correspondence on minimal surfaces.
RE was supported by DURSI 2009 SGR 168, MEC
FPA 2007-66665-C02 and CPAN CSD2007-00042 Consolider-Ingenio 2010. TH
was supported by the Carlsberg foundation. VN was supported by an
Individual Marie Curie Intra-European Fellowship and by
ANR-05-BLAN-0079-02 and MRTN-CT-2004-503369, and CNRS PICS {\#} 3059,
3747 and 4172.


\begin{appendix}

\section{General equations for stationary black $S^3$-folds}
\label{s3}

In this appendix we consider the general equations for stationary blackfolds
with horizon topology
\beq
\label{s3aa}
S^3 \times s^{n+1}
~.
\eeq
We embed the spatial part of the blackfold worldvolume, $\BB_3$, in a
four-dimensional subspace of $\R^{D-1}$ whose metric we parametrize as
\beq
\label{s3ab}
d\rho^2+\rho^2 (d \theta^2 +\sin^2\theta d \phi^2+\cos^2\theta d \psi^2)
~, ~~ 0\leq \theta \leq \frac{\pi}{2}~, ~~ 0\leq \phi,\psi < 2\pi
~.
\eeq
$\BB_3$ is wrapped around the three-sphere with coordinates $(\theta,\phi,\psi)$
and its embedding in $\R^4$ is provided by the radial scalar $\rho=R(\theta)$.
Allowing for rigid rotation along the Killing directions associated with the angles
$\phi$ and $\psi$, the velocity field $u$ is oriented along the Killing vector
\beq
\label{s3ac}
{\bf k}=\frac{\d}{\d t}+\Omega_1 \frac{\d}{\d \phi}+\Omega_2 \frac{\d}{\d \psi}
~.
\eeq
Consequently,
\beq
\label{s3ad}
R_0=1~, ~~ V(\theta)^2=R(\theta)^2(\Omega_1^2 \sin^2\theta+\Omega_2^2 \cos^2\theta)
~.
\eeq
The thickness $r_0$ is also a function of $\theta$
\beq
\label{s3ae}
r_0(\theta)=\frac{n}{2\kappa}
\sqrt{1-R(\theta)^2(\Omega_1^2 \sin^2\theta+\Omega_2^2 \cos^2\theta)}
~.
\eeq

With these data the stationary blackfold action \eqref{BFact} becomes
\beq
\label{s3af}
I(R)=\int_0^{\frac{\pi}{2}} d \theta \, \sin\theta \cos\theta
R(\theta)^2 \sqrt{R(\theta)^2+R'(\theta)^2}
\Big( 1- R(\theta)^2(\Omega_1^2 \sin^2\theta+\Omega_2^2 \cos^2\theta)
\Big)^{\frac{n}{2}}
~.
\eeq
Varying with respect to $R(\theta)$ we obtain the equation of motion
\bea
\label{s3ag}
&&-(n-2) \Omega_+\Omega_- R^2 R'(R^2+{R'}^2)
-2 \cos(2\theta) \Big( -1+(\Omega_+^2+\Omega_-^2)R^2\Big)(R^2+{R'}^2) R'
\nonumber\\
&&+ \sin(2\theta) R\Big( -4 {R'}^2+R(R''+R(-3+(\Omega_+^2+\Omega_-^2)
((n+3)R^2+(n+4){R'}^2-RR'')))\Big)
\nonumber\\
&&+(n+2) \Omega_+\Omega_- \cos(4\theta) R^2 R'(R^2+{R'}^2)
\nonumber\\
&&-\Omega_+\Omega_- \sin(4\theta) R^3\Big( (n+3) R^2+(n+4) {R'}^2-RR''\Big)=0
~.
\eea
We have defined
\beq
\label{s3ai}
\Omega_\pm=\frac{1}{2}(\Omega_1\pm \Omega_2)
~.
\eeq

Setting $\Omega_-=0$ and $R=$constant in \eqref{s3ag} we recover
easily the round sphere solution of Subsection \ref{singlesphere} with
\beq
\label{s3aj}
R=\sqrt{\frac{3}{n+3}}\frac{1}{\Omega_+}
\eeq
which is in exact agreement with eq.\ \eqref{oddSai} for $p=3$.

It is an interesting problem to find inhomogeneous solutions of eq.\ \eqref{s3ag}
with $\Omega_\pm \neq 0$. It appears that numerical analysis will be needed
for this task. We will not pursue this task here, but will exhibit a perturbative inhomogeneous
solution for finite $\Omega_+$ and small $\Omega_-$.

Setting
\beq
\label{s3ak}
R(\theta)=\sqrt{\frac{3}{n+3}}\frac{1}{\Omega_+}+\Omega_- r(\theta)~, ~~
\Omega_-\ll \Omega_+
\eeq
we expand the differential equation \eqref{s3ag} to leading order in $\Omega_-$,
solve for $r(\theta)$ and find the following inhomogeneous solution for generic $n$
\beq
\label{s3al}
R(\theta)=\sqrt{\frac{3}{n+3}}\frac{1}{\Omega_+}
-\frac{3\sqrt{3(n+3)}}{n-9} \frac{\Omega_-}{\Omega_+^2} \cos(2\theta)
+ \OO(\Omega_-^2)
~.
\eeq

The values $n=1$ and $n=9$ appear to be special. For $n=1$ we find
a one-parameter family of regular solutions at this order
\beq
\label{s3am}
R(\theta)=\frac{\sqrt{3}}{2}\frac{1}{\Omega_+}
+\frac{3\sqrt{3}}{4}\frac{\Omega_-}{\Omega_+^2}\cos(2\theta)
+c\Omega_- \Big(1+3 \cos(4\theta)\Big)+\OO(\Omega_-^2)
\eeq
where $c$ is a free parameter whose physical meaning is unclear.
For $n=9$ we do not find a regular solution. It would be interesting to clarify
these peculiarities and obtain a better understanding of these solutions
numerically beyond perturbation theory.

\section{Physical magnitudes for even-ball blackfolds and ultraspinning
MP black holes}

\label{app:ballsMP}

\subsection{Even-balls}

Inserting the velocity field
\beq
V^2=\sum_{i=1}^k\Omega_i^2r_i^2
\eeq
and \eqref{ballaf} into \eqref{mass}, \eqref{angmom}, \eqref{Sbf} we find
\begin{subequations}
\beq
M=\frac{r_+^n \Omega_{(n+1)} }{16\pi G}(2\pi)^k \int_{\EE}
\prod_{\ell=1}^k dr_\ell ~r_\ell
\left(n+1-\sum_i \Omega_i^2 r_i^2\right)
\left(1-\sum_i \Omega_i^2 r_i^2\right)^{\frac{n}{2}-1}
~,
\eeq
\beq
J_i=\frac{nr_+^n \Omega_{(n+1)} }{16\pi G}(2\pi)^k \int_{\EE}
\prod_{\ell=1}^k dr_\ell ~r_\ell r_i
\left(1-\sum_\ell \Omega_\ell^2 r_\ell^2\right)^{\frac{n}{2}-1}
\Omega_i r_i
~,
\eeq
\beq
S=\frac{r_+^{n+1}\Omega_{(n+1)}}{4G} (2\pi)^k  \int_{\EE}
\prod_{\ell=1}^k dr_\ell ~r_\ell \left(1-\sum_\ell \Omega_\ell^2 r_\ell^2\right)^{\frac{n}{2}}
~.
\eeq
\end{subequations}

We have denoted the range of the $k$-dimensional integration as
\beq
\EE=\{ \vec r ~:~ r_i \geq 0~, ~\sum_\ell \Omega_\ell^2 r_\ell^2 \leq 1 \}
~.
\eeq
By changing variables to
\beq
x_i=\Omega_i r_i
\eeq
we convert the range of integration to a piece of a unit $k$-ball, which we will denote
as $\EE_1$ (it would be a unit $k$-ball if $r_i$ were allowed to be both positive
and negative). Thus, we get
\begin{subequations}
\beq
M=\frac{r_+^n \Omega_{(n+1)} }{16\pi G}
\prod_{\ell=1}^k \left(\frac{2\pi}{\Omega_\ell^2}\right)
\int_{\EE_1} \prod_{\ell=1}^k dx_\ell ~x_\ell
\left(n+1-\sum_\ell x_\ell^2\right)
\left(1-\sum_\ell x_\ell^2\right)^{\frac{n}{2}-1}
~,
\eeq
\beq
J_i=\frac{nr_+^n \Omega_{(n+1)} }{16\pi G}
\prod_{\ell=1}^k \left(\frac{2\pi}{\Omega_\ell^2}\right)\Omega_i^{-1}
\int_{\EE_1} \prod_{\ell=1}^k dx_\ell ~x_\ell x_i^2
\left(1-\sum_\ell x_\ell^2\right)^{\frac{n}{2}-1}
~,
\eeq
\beq
S=\frac{\Omega_{(n+1)} }{4G}r_+^{n+1}
\prod_{\ell=1}^k \left(\frac{2\pi}{\Omega_\ell^2}\right)
\int_{\EE_1} \prod_{\ell=1}^k dx_\ell ~x_\ell
\left(1-\sum_\ell x_\ell^2\right)^{\frac{n}{2}}
~.
\eeq
\end{subequations}

To compute these integrals we set
\beq
x_i=r \mu_i
~,
\eeq
where $\mu_i$ are $k$ director cosines for $S^{k-1}$ spheres and $r\in [0,1]$.
Then, integrals of the form
\beq
\int_{\EE_1} \prod_{\ell=1}^k dx_\ell ~x_\ell
f\left(\sum_\ell x_\ell^2\right)
\nonumber
\eeq
become
\beq
\int_0^1 dr \, r^{2k-1} f(r^2) ~\int ' d\Omega_{(k-1)}  \prod_{j=1}^k \mu_j
~,
\nonumber
\eeq
where the prime over the sphere integral stands to remind us that
we should not forget to impose the constraint $r_i\geq 0$.

Notice that we can decompose the metric of a $(2k-1)$-sphere as
\beq
d\Omega_{(2k-1)}  ^2=d\Omega_{(k-1)} ^2+\sum_{i=1}^k \mu_i^2 d\psi_i^2
~.
\eeq
We can deduce from here that
\beq
\label{intprime}
\int ' d\Omega_{(k-1)}  \prod_{\ell=1}^k \mu_\ell=
\frac{\Omega_{(2k-1)} }{(2\pi)^k}
~.
\eeq
To determine $J_i$
we use the following formula based on symmetry
\beq
\int_{\EE_1} \prod_{\ell=1}^k dx_\ell ~x_\ell x_i^2 f\left(\sum_j x_j^2 \right)=
\frac{1}{k} \int_{\EE_1} \prod_{\ell=1}^k dx_\ell~ x_\ell \left( \sum_j x_j^2 \right)
f\left(\sum_j x_j^2 \right)
~.
\eeq
With this and \eqref{intprime} we obtain the simple expressions
\eqref{ballaea}-\eqref{ballaec} for the mass, angular momenta and area.

\subsection{Ultraspinning MP black holes}

Consider a Myers-Perry black hole in $D$ dimensions, with $k$ non-vanishing
angular momenta. The solution is parametrized in terms of a mass
parameter $\mu$ and $k$ rotation parameters $a_i$. The mass, angular
momenta and angular velocities are
\begin{subequations}
\beq\label{mpmass}
M=\frac{D-2}{16\pi G}\Omega_{(D-2)}\mu\,,
\eeq
\beq
J_i=\frac{2}{D-2}M a_i\,,
\eeq
\beq
\Omega_i=\frac{a_i}{r_+^2+a_i^2}\,.
\eeq
\end{subequations}
The properties of the horizon, such as the radius $r_+$, surface gravity
and area cannot be easily found for generic parameters, but they
simplify dramatically in the ultraspinning limit in which \cite{Emparan:2003sy}
\beq
a_i\gg \mu^{1/(D-3)}
\eeq
(we assume $a_i>0$ without loss of generality). Then
\beq
r_+ \to \left(\frac{\mu}{\prod_{i=1}^k a_i^2}\right)^{1/(D-2k-3)}\,,
\eeq
and
\beq
\kappa\to \frac{D-2k-3}{2r_+}\,,
\eeq
\beq
S\to\frac{\Omega_{(D-2)}}{4G}r_+^{D-2k-2}\prod_{i=1}^k a_i^2\,.
\eeq
Then, substituting
\beq\label{nsphe}
\Omega_{(D-2)}=\frac{2\pi^{\frac{D-1}{2}}}{\Gamma(\frac{D-1}{2})}
\eeq
into \eqref{mpmass} we obtain
\beq
M\to \frac{r_+^{D-2k-3}}{8G}\frac{(D-2)\pi^\frac{D-3}{2}}{\Gamma\left(\frac{D-1}{2}\right)}
\prod_{i=1}^k a_i^2\,.
\eeq
Also in this limit,
\beq
\Omega_i\to \frac{1}{a_i}\,.
\eeq

The horizon has a round sphere
$S^{D-2k-2}$ in directions transverse to the rotation planes with radius
\beq
R(S^{D-2k-2})\to r_+\cos\theta
\eeq
where $\theta$ is a polar angle (more precisely, if the rotating
directions are parametrized using direction cosines $\mu_i$, then
$\sum_{i=1}^k\mu_i^2=\sin^2\theta$).

If we now set $n=D-2k-3$ and identify the angular velocities and surface
gravity of this solution with those of a $2k$-ball blackfold (so $r_+$
is indeed the same in both constructions), then the
expressions for the mass, angular momenta, and entropy
\eqref{ballaea}-\eqref{ballaec}, and radius of the
transverse spheres \eqref{ballai} are exactly reproduced.

\section{Matched asymptotic expansions for black tori}
\label{tori}

In this appendix we present the main steps of a matched asymptotic expansion
(MAE) that determines the full metric to leading order in the expansion parameter $r_0/R$
of rotating black tori in arbitrary spacetime dimension. This analysis generalizes the
corresponding solutions of perturbative black rings in \cite{Emparan:2007wm}. It
provides perturbative black hole solutions with horizon topology $\T^p \times s^{n+1}$,
for generic $p,n$ (satisfying $p \leq n+2$) and confirms that the horizon remains
regular when the stationary blackfold equations are satisfied.

\subsection{A note on coordinate systems}
\label{coord}

In order to study the effects of wrapping a black $p$-brane around an orthogonal
$p$-dimensional torus $\T^p$ with radii $R_i$ ($i=1,\ldots,p$)%
\footnote{Note we have a slight change of notation $R_a \rightarrow R_i$ compared to
the general discussion in Subsection \ref{prododd}.} it is convenient
to introduce coordinates appropriately adapted to the torus. The asymptotic
spacetime is $D$-dimensional Minkowski spacetime with $D=n+p+3$. We
write its metric in the following coordinate system
\beq
\label{coordaa}
ds^2_D=-dt^2+\sum_{i=1}^p\left( dr_i^2+r_i^2 d\psi_i^2\right)
+\sum_{k=1}^{n+2-p}dx_k^2
~.
\eeq
We envision wrapping the black $p$-brane around the surface parametrized
by the toric angles $\psi_i$ at $r_i=R_i$ and $x_k=0$.

It is useful to define the radial coordinate $r$ and the director cosines $\mu_a$,
$a=1,\ldots, n+2$, such that
\beq
\label{coordab}
r_i=R_i+r\mu_i~, ~~ i=1,\ldots, p~, ~~~~
x_k=r\mu_{p+k}~, ~~ k=1,\ldots, n+2-p
\eeq
with
\beq
\label{coordac}
\sum_{a=1}^{n+2}\mu_a^2=1~, ~~ \sum_{a=1}^{n+2}d\mu_a^2=d\Omega_{(n+1)}^2
~.
\eeq
In the near-object limit, where $R_i\to \infty$ with $r$, $\mu_a$, $z_i=R_i\psi_i$ fixed,
the Minkowski metric becomes
\beq
\label{coordad}
ds^2_D\simeq -dt^2+dr^2+r^2d\Omega_{(n+1)}^2
+\sum_{i=1}^p \left( 1+2 \frac{r\mu_i}{R_i}\right)dz_i^2+\OO\left( \frac{r^2}{R_i^2}\right)
~.
\eeq

It is also convenient to introduce another coordinate system, call it $r$-adapted
coordinate system, for which the new $r$-coordinate defines scalar equipotential
surfaces, $i.e.$ $\nabla^2r^{-n}=0$, to leading order in $1/R_i$. The leading order
transformation that defines this system is (for $i=1,\ldots,p$, $k=1,\ldots, n+2-p$)
\beq
\label{coordae}
r\to r-\frac{r^2}{2n}\sum_{j=1}^p \frac{\mu_j}{R_j}~, ~~
\mu_i \to \mu_i-\frac{r}{2n} \left( \mu_i \sum_{j=1}^p \frac{\mu_j}{R_j}-\frac{1}{R_i}\right)
~,~~
\mu_{k+p}\to \mu_{k+p}\left( 1-\frac{r}{2n}\sum_{j=1}^p \frac{\mu_j}{R_j}\right)
~.
\eeq
In these coordinates the Minkowski metric becomes
\beq
\label{coordaf}
ds^2_D\simeq-dt^2+ \left( 1-\sum_{i=1}^p \frac{2r \mu_i}{nR_i}\right)
\left(dr^2+r^2 d\Omega_{(n+1)} ^2\right)+
\sum_{i=1}^p \left(1+\frac{2r \mu_i}{R_i}\right)dz_i^2+\OO\left(\frac{r^2}{R_i^2}\right)
~.
\eeq

\subsection{The far-zone}
\label{far}

The first step of a MAE is to find a solution of a thin black $p$-torus, with $s^{n+1}$
radius $r_0\ll R_i$, in the far-zone, $r\gg r_0$, where the linearized approximation
to gravity is valid.

We solve the linearized Einstein equations in transverse gauge,
\beq
\label{faraa}
\Box \bar h_{\mu\nu}=-16\pi G T_{\mu\nu}
\eeq
with $\bar h_{\mu\nu}=h_{\mu\nu}-\frac{1}{2}h^\rho_\rho g_{\mu\nu}$ and
$\nabla_\mu \bar h^{\mu\nu}=0$. $h_{\mu\nu}$ is the linear perturbation of the
background metric $g_{\mu\nu}$. The source should be chosen to have
support on a $p$-torus at $r_i=R_i$, $x_k=0$.
Locally it should reproduce the stress-energy tensor of a boosted black $p$-brane
solution. Choosing the critical boost for which the stationary blackfold equations,
$T_{ii}=0$, are satisfied we obtain
\begin{subequations}
\label{farab}
\beq
\label{faraba}
T_{tt}=\frac{\Omega_{(n+1)}}{16\pi G}(n+p+1) r_0^n \prod_{i=1}^p \delta (r_i-R_i)\delta(x_k)
~,
\eeq
\beq
\label{farabb}
T_{ti}=\frac{\Omega_{(n+1)}}{16\pi G}\sqrt{n+p}\, r_0^n \prod_{i=1}^p \delta (r_i-R_i)\delta(x_k)
~,
\eeq
\beq
\label{farabc}
T_{ii}=0~, ~~
T_{ij}=\frac{\Omega_{(n+1)}}{16\pi G} r_0^n \prod_{i=1}^p \delta (r_i-R_i)\delta(x_k)
~, ~~ i\neq j
~.
\eeq
\end{subequations}
To integrate the equation \eqref{faraa} we use the Green's function in $\R^{D-1}$
\beq
\label{farac}
G(x,x')=-\frac{1}{(D-3)\Omega_{(D-2)}}\frac{1}{|x-x'|^{D-3}}
~.
\eeq
Notice that the stationary blackfold equations, $T_{ii}=0$, which are equivalent to the
conservation of the stress-energy tensor, are necessarily satisfied in the transverse
gauge.

It is now possible to obtain explicit integral expressions for the metric perturbation.
Here we are mainly interested in the form of these expressions in the near-object
limit that was defined below eq.\ \eqref{coordac}. The form of the metric in this
regime will provide boundary conditions for the near-zone metric of the next
subsection. In the $r$-adapted coordinate system, and in Schwarzschild gauge
\beq
\label{farad}
r\to r-\frac{1}{2n}\frac{r_0^n}{r^{n-1}}
~,
\eeq
the far-zone metric perturbation reads
\begin{subequations}
\label{farae}
\beq
\label{faraea}
g_{tt} = -1 + \frac{n+p}{n} \frac{r_0^n}{r^n}
~,
\eeq
\begin{equation}
\label{faraeb}
g_{ti} = \frac{\sqrt{n+p}}{n} \frac{r_0^n}{r^n}
\left( 1 + \frac{ r \mu_i}{R_i}  \right)
~,
\end{equation}
\begin{equation}
\label{faraec}
g_{i \neq j} = \frac{1}{n} \frac{r_0^n}{r^n}
\left( 1 + \frac{ r \mu_i}{R_i} + \frac{ r \mu_j}{R_j}   \right)
~,
\end{equation}
\begin{equation}
\label{faraed}
g_{i i} = 1 + \frac{1}{n} \frac{r_0^n}{r^n} \left( 1 + \frac{ r \mu_i}{R_i} \right)
 +  \frac{2 r \mu_i}{R_i}
~,
\end{equation}
\begin{equation}
\label{faraee}
g_{rr} = 1 + \frac{r_0^n}{r^n} \left( 1 - \frac{2n-1}{n^2} \sum_{i=1}^p
\frac{ r \mu_i}{R_i} \right) - \frac{2}{n} \sum_{i=1}^p \frac{r \mu_i}{ R_i}
~,
\end{equation}
\begin{equation}
\label{faraef}
g_{\Omega \Omega} = 1 + \frac{1}{n^2}
\frac{r_0^n}{r^n} \sum_{i=1}^p
\frac{ r \mu_i}{R_i}  - \frac{2}{n} \sum_{i=1}^p \frac{ r \mu_i}{R_i}
~.
\end{equation}
\end{subequations}

\vspace{-0.6cm}
\subsection{The near-zone}
\label{near}

The next step requires solving the Einstein equations in the near-zone
region, $r\ll R_i$. Here we are perturbing around the critically boosted
black $p$-brane metric. Following the arguments of \cite{Emparan:2007wm},
we expect the perturbations to take the form
\begin{equation}
\label{nearaa}
h_{\mu \nu} = \sum_{i=1}^p \frac{\mu_i}{R_i} a^{(i)}_{\mu \nu} (r)
\end{equation}
Since the perturbations for each $i = 1\ldots p $ will decouple to leading order in
this expansion, we may consider for simplicity a given $i=k$. Hence, we perturb
the critically boosted black $p$-brane in the following way
\begin{subequations}
\label{gmunu}
\begin{equation}
\label{gtta}
g_{tt} = - 1 + \frac{n+p}{n} \frac{r_0^{n}}{r^{n}} +
\frac{\mu_k} {R_k} a (r)\,,
\end{equation}
\begin{equation}
g_{ti} = -\frac{\sqrt{n+p}}{n} \left[ \frac{r_0^{n}}{r^{n}} +
\frac{\mu_k}{R_k} b_i(r) \right]\,,
\end{equation}
\begin{equation}
g_{ij} = \delta_{ij}  + \frac{1}{n} \frac{r_0^{n}}{r^{n}} + \frac{\mu_k}{R_k}
c_{ij} (r)\,,
\end{equation}
\begin{equation}\label{grr}
g_{rr} = \left( 1- \frac{r_0^{n}}{r^{n}} \right)^{-1} \left[ 1
+ \frac{\mu_k}{R_k} f(r) \right]\,,
\end{equation}
\begin{equation}
\label{gijg}
g_{\Omega \Omega} = 1
+ \frac{\mu_k}{R_k} g(r) \,.
\end{equation}
\end{subequations}
By definition $g_{\Omega \Omega}$ multiplies the metric
$r^2 \sum_{j=1}^{n+2} d \mu_j^2$ of  a $S^{n+1}$ of radius $r$.
In the above formulae we have omitted for brevity a label $k$ on the
functions $a$, $b_i$, $c_{ij}$, $f$ and $g$.

With this ansatz, the location of the horizon will remain at $r=r_0$ if
the perturbations are finite there. This fixes partially the choice of
radial coordinate, but there still remains some gauge freedom under the
transformation
\begin{subequations}
\label{gauge}
\beqa
r&\to&r+ \gamma(r)\,\frac{r_0}{R_k}\mu_k \,,\\
\mu_i &\to& \mu_i + \beta(r)\,\frac{r_0}{R_k} ( \mu_i \mu_k - \delta_{ik} )
\eeqa
\end{subequations}
with
\beq\label{bgauge}
\beta'(r)=\frac{\gamma(r)}{r^2\left(1- \frac{r_0^{n}}{r^{n}}\right)}
~.
\eeq
With this change of coordinates the metric retains its form to leading order in $1/R_k$.
The condition that the horizon stays at $r=r_0$, $i.e.$ $r_0\to r_0 +O(1/R_k^2)$, is
\beq
\label{gammao}
\gamma(r_0)=0
~.
\eeq
In addition, there is an $r$-independent dipole gauge transformation,
\beq\label{zeromode}
\mu_i \to  \mu_i + \beta_0\,\frac{r_0}{R_k} ( \mu_i \mu_k - \delta_{ik} )
\eeq
with constant $\beta_0$ that leaves everything unchanged except for $g(r)$ which
is redefined by an additive constant
\beq
g(r)\to g(r)+2 r_0 \beta_0
~.
\eeq

Let us summarize how many undetermined functions we have.
Based on symmetry considerations we expect (again for fixed $k$) that
\begin{equation}
\label{nearab}
b_{k} = b  \spa b_{i \neq k} = \widetilde b \spa
c_{kk} = c \spa c_{k , i\neq k } = \widetilde c \spa c_{i\neq k,j\neq k} = \widehat c
~.
\end{equation}
The unknown functions are therefore $a, b, \widetilde b, c,\widetilde c, \widehat c,f,g$.
From these we can form the gauge-invariant combinations
\begin{subequations}
\label{invar}
\beqa
\label{Ainvt}
{\sf A}(r)&=&a(r)-(n+p)\,c (r)\,,\\
{\sf B}(r)&=&b(r)-n \,c(r)\,,\label{Binvt}\\
{\sf F}(r)&=&f(r)+2 r_0\left(\frac{r^{n+1}}{r_0^{n+1}}\;c(r)\right)'
-\frac{n}{\left(1-\frac{r_0^{n}}{r^{n}}\right)}\;c(r)\,,
\label{Finvt}\\
{\sf G}'(r)&=&g'(r)+2 \frac{r_0}{r}\left(\frac{r^{n+1}}{r_0^{n+1}}\;c(r)\right)'
+\frac{2}{r\left(1- \frac{r_0^{n}}{r^{n}}\right)}\;c(r)\label{Gpinvt}
\, \label{extrainvt}\\
\widetilde{\sf B} (r) &=& \widetilde b(r) - n c (r)
\spa \widetilde {\sf H} (r) = \widetilde c(r) - c (r)
\spa  \widehat {\sf H} (r) = \widehat c(r) - c (r)
~.
\eeqa
\end{subequations}
The first four functions are the same as in the black ring case with $p=1$
\cite{Emparan:2007wm}. For $p>1$ there are three additional functions.

The far-zone analysis (see \eqref{farae}) imposes the following boundary conditions
on these functions
\begin{subequations}
\label{uas}
\beqa
{\sf A}(r)&=&-2(n+p)\, r-\frac{n+p}{n}\frac{r_0^n}{r^{n-1}}+\OO\left(r^{-n-1}\right)\,,\\
{\sf B} (r) &=&-2n\, r+\OO\left(r^{-n-1}\right)\, , \\
{\sf F} (r) &=& 4 (n +2) \frac{r^{n+1}}{r_0^n} + \frac{2}{n} (1-n^2) r +
\OO\left(r^{-(n-1)}\right) \, , \\
{\sf G}' (r) &=&  4 (n +2) \frac{r^{n}}{r_0^n} + \frac{2}{n} (1+ 2n) +
\OO\left(r^{-n}\right)\
\, , \\
\widetilde{\sf B} (r) &=& -2nr- \frac{r_0^n}{r^{n-1}}+\OO\left(r^{-n-1}\right)\, , \\
\widetilde {\sf H} (r) &=& - 2r+O\left(r^{-n-1}\right)\, , \\
\widehat {\sf H} (r) &=& -2r- \frac{1}{n} \frac{r_0^n}{r^{n-1}}+O\left(r^{-n-1}\right)
~.
\eeqa
\end{subequations}

We can now proceed to determine the gauge invariant functions
by solving the Einstein equations. The computation simplifies
considerably by noting that the boundary conditions \eqref{uas}
and the form of the Einstein equations depend non-trivially only
on $n$, and that the $p$-dependence can be incorporated easily
with appropriate rescalings.

From the $R_{r\theta}=0$ equation we can determine ${\sf F}(r)$ in terms
of the other functions. $\theta$ is one of the angles parametrizing the director
cosine $\mu_k$ that appears in the perturbation \eqref{gmunu}.
Then, from linear combinations of the equations associated to
$R_{tt}, R_{rr}, R_{\Omega\Omega}$ we can solve for the function
${\sf G}'(r)$ in terms of the functions ${\sf A}(r)$, ${\sf B}(r)$, ${\sf B}(r)$,
$\widetilde {\sf H}(r)$, $\widehat {\sf H}(r)$ and their first derivatives.
The remaining equations simplify considerably if we define and substitute
for the following auxiliary functions
\beq
\label{genpnaa}
\KK_1=2\widetilde{\sf H}-\widehat {\sf H}~, ~~
\KK_2=-n\widetilde {\sf H}+\widetilde{\sf B}~, ~~
\KK_3=\widetilde{\sf B}-n\widehat{\sf H}
~,
\eeq
\beq
\label{genpnab}
\MM={\sf B}-\KK_2~, ~~
\SS={\sf B}-\KK_3~, ~~
\CC=(n+1)\left(-\frac{n}{n+p} {\sf A}+2{\sf B}\right)
~.
\eeq
The functions $\KK_1$, $\MM$ and $\CC$ obey the same decoupled
$p$-independent differential equation
\beq
\label{genpnac}
-(n+1)r^{n-1} {\sf Y}+\left( (n+1)r^n-r_0^n\right){\sf Y}'
+r\left(r^n-r_0^n\right){\sf Y}''=0
~.
\eeq
$\SS$ and ${\sf B}$ obey the differential equation
\beq
\label{genpnad}
-(n+1)r^{n-1} {\sf X}+(n+1)\left( r^n-r_0^n\right){\sf X}'
+r\left(r^n-r_0^n\right){\sf X}''+n(n+1)r_0^n \KK'_1=0
~.
\eeq

From the first set of equations \eqref{genpnac} and the boundary
conditions \eqref{uas} we deduce the simple linear relations
\beq
\label{genpnae}
\CC(r)=(n+1) \MM(r)~, ~ ~ \MM(r)=n \KK_1(r)
~.
\eeq
From \eqref{genpnad} and the boundary conditions we deduce that
\beq
\label{genpnaf}
\SS(r)={\sf B}(r) ~~ \Leftrightarrow ~~ \KK_3(r)=0
~.
\eeq
These equations are enough to determine the functions $\widetilde{\sf B}(r)$,
$\widetilde {\sf H}(r)$, $\widehat {\sf H}(r)$ in terms of ${\sf A}(r)$, and
${\sf B}(r)$. We find
\beq
\label{genpnag}
\widetilde {\sf H}(r)=\frac{1}{n} {\sf B}(r)~, ~ ~
\widetilde{\sf B}(r)=\frac{n}{n+p} {\sf A}(r)~, ~ ~
\widehat{\sf H}(r)=\frac{1}{n+p} {\sf A}(r)
~.
\eeq

Finally, we differentiate \eqref{genpnac} (applied to $\KK_1$) once
with respect to $r$ and substitute for the derivatives of $\KK_1$
the expression that follows from \eqref{genpnad} in terms of ${\sf B}$.
Then we get for ${\sf B}$ a fourth order master equation, the same
equation that appeared for $p=1$ in \cite{Emparan:2007wm}
(eq.\ (6.16) in that paper). The master equation can be solved by a
linear combination of hypergeometric functions (see
\cite{Emparan:2007wm} for details). Using the above equations the
function ${\sf A(r)}$ can be expressed in terms of ${\sf B}(r)$ and its first,
second and third derivatives (the relation is given by eq.\ (6.15) in
\cite{Emparan:2007wm} by replacing ${\sf B}\to \frac{n+1}{n+p}{\sf A}$ and
${\sf A} \to {\sf B}$). So, once we determine ${\sf B}$ all remaining
functions are determined.

\subsection{Matching and the complete solution}
\label{complete}

The boundary conditions \eqref{uas} provide the matching between the
far-zone and near-zone expansions. We can use them to solve the master
equation for ${\sf B}$, then from ${\sf B}$ we can determine the other gauge
invariant functions and finally we can fix the gauge to obtain the complete metric.
The details of this exercise proceed {\it mutatis mutandis} to the black
ring case that was analyzed in \cite{Emparan:2007wm}. We refer the reader
to that paper for further details. We remark here that since the perturbations are
of dipole type the physical quantities do not receive modifications to order $1/R_i$.
Moreover, as observed
  for the black ring family $S^1 \times  s^{n+1}$ in Ref.~\cite{Emparan:2007wm},
the hypergeometric functions entering in the solution presented above
simplify drastically for $n=1$. As a consequence, the first-order
corrected metric is not only simpler for the five-dimensional black
ring, but also for the horizon topologies
$\T^2 \times s^2$ and $\T^3 \times s^2$ in six and seven dimensions respectively.

\end{appendix}

\addcontentsline{toc}{section}{References}

\begin{thebibliography}{10}

\bibitem{Emparan:2009cs}
R.~Emparan, T.~Harmark, V.~Niarchos, and N.~A. Obers, {\it World-volume
  effective theory for higher-dimensional black holes. ({B}lackfolds)},  {\em
  Phys. Rev. Lett.} {\bf 102} (2009) 191301,
  [\href{http://xxx.lanl.gov/abs/0902.0427}{{\tt 0902.0427}}].

\bibitem{Emparan:2009at}
R.~Emparan, T.~Harmark, V.~Niarchos and N.~A.~Obers,
  {\it {Essentials of Blackfold Dynamics}},
  {\em JHEP} {\bf 1003}, 063 (2010)
  [\href{http://xxx.lanl.gov/abs/0910.1601}{{\tt 0910.1601}}].

\bibitem{Emparan:2007wm}
R.~Emparan, T.~Harmark, V.~Niarchos, N.~A. Obers, and M.~J. Rodriguez, {\it
  {The Phase Structure of Higher-Dimensional Black Rings and Black Holes}},
  {\em JHEP} {\bf 10} (2007) 110,
  [\href{http://xxx.lanl.gov/abs/0708.2181}{{\tt 0708.2181}}].

\bibitem{Obers:2008pj}
N.~A. Obers, {\it {Black Holes in Higher-Dimensional Gravity}},  {\em Lect.
  Notes Phys.} {\bf 769} (2009) 211--258,
  [\href{http://xxx.lanl.gov/abs/0802.0519}{{\tt 0802.0519}}].

\bibitem{Niarchos:2008jc}
V.~Niarchos, {\it {Phases of Higher Dimensional Black Holes}},  {\em Mod. Phys.
  Lett.} {\bf A23} (2008) 2625--2643,
  [\href{http://xxx.lanl.gov/abs/0808.2776}{{\tt 0808.2776}}].

\bibitem{Emparan:2008eg}
R.~Emparan and H.~S. Reall, {\it {Black Holes in Higher Dimensions}},  {\em
  Living Rev. Rel.} {\bf 11} (2008) 6,
  [\href{http://xxx.lanl.gov/abs/0801.3471}{{\tt 0801.3471}}].

\bibitem{Caldarelli:2008pz}
M.~M. Caldarelli, R.~Emparan, and M.~J. Rodriguez, {\it {Black Rings in
  (Anti)-de{S}itter space}},  {\em JHEP} {\bf 11} (2008) 011,
  [\href{http://xxx.lanl.gov/abs/0806.1954}{{\tt 0806.1954}}].

\bibitem{Camps:2008hb}
J.~Camps, R.~Emparan, P.~Figueras, S.~Giusto, and A.~Saxena, {\it {Black Rings
  in {Taub-NUT} and {D0-D6} interactions}},  {\em JHEP} {\bf 02} (2009) 021,
  [\href{http://xxx.lanl.gov/abs/0811.2088}{{\tt 0811.2088}}].

\bibitem{LeWitt:2009qx}
J.~Le Witt and S.~F.~Ross,
  {\it {Black holes and black strings in plane waves}},
  [\href{http://xxx.lanl.gov/abs/0910.4332}{{\tt 0910.4332}}].

\bibitem{BlancoPillado:2007iz}
J.~J. Blanco-Pillado, R.~Emparan, and A.~Iglesias, {\it {Fundamental Plasmid
  Strings and Black Rings}},  {\em JHEP} {\bf 01} (2008) 014,
  [\href{http://xxx.lanl.gov/abs/0712.0611}{{\tt 0712.0611}}].

\bibitem{Hollands:2006rj}
S.~Hollands, A.~Ishibashi, and R.~M. Wald, {\it A higher dimensional stationary
  rotating black hole must be axisymmetric},
  \href{http://xxx.lanl.gov/abs/gr-qc/0605106}{{\tt gr-qc/0605106}}.

\bibitem{Moncrief:2008mr}
  V.~Moncrief and J.~Isenberg, {\it Symmetries of Higher Dimensional Black Holes},
  {\em Class. Quant. Grav.}  {\bf 25} (2008) 195015,
  [\href{http://xxx.lanl.gov/abs/0805.1451}{{\tt 0805.1451}}].

\bibitem{Reall:2002bh}
H.~S. Reall, {\it Higher dimensional black holes and supersymmetry},
  \href{http://xxx.lanl.gov/abs/hep-th/0211290}{{\tt hep-th/0211290}}.

\bibitem{Kudoh:2004hs}
H.~Kudoh and T.~Wiseman, {\it Connecting black holes and black strings},  {\em
  Phys. Rev. Lett.} {\bf 94} (2005) 161102,
  [\href{http://xxx.lanl.gov/abs/hep-th/0409111}{{\tt hep-th/0409111}}].

\bibitem{Kol:2004ww}
B.~Kol, {\it The phase transition between caged black holes and black strings:
  {A} review},  {\em Phys. Rept.} {\bf 422} (2006) 119--165,
  [\href{http://xxx.lanl.gov/abs/hep-th/0411240}{{\tt hep-th/0411240}}].

\bibitem{Harmark:2007md}
T.~Harmark, V.~Niarchos, and N.~A. Obers, {\it Instabilities of black strings
  and branes},  {\em Class. Quant. Grav.} {\bf 24} (2007) R1--R90,
  [\href{http://xxx.lanl.gov/abs/hep-th/0701022}{{\tt hep-th/0701022}}].

\bibitem{Colding}
T. Colding and W.P. Minicozzi, {\it Minimal surfaces}, Courant Lecture Notes
in Mathematics, 4. New York University, Courant Institute of Mathematical Sciences, New York, 1999.

\bibitem{Gibbons:2002bh}
G.~W. Gibbons, D.~Ida, and T.~Shiromizu, {\it Uniqueness and non-uniqueness of
  static vacuum black holes in higher dimensions},  {\em Prog. Theor. Phys.
  Suppl.} {\bf 148} (2003) 284--290,
  [\href{http://xxx.lanl.gov/abs/gr-qc/0203004}{{\tt gr-qc/0203004}}].

\bibitem{Carter:2000wv}
B.~Carter, {\it Essentials of classical brane dynamics},  {\em Int. J. Theor.
  Phys.} {\bf 40} (2001) 2099--2130,
  [\href{http://xxx.lanl.gov/abs/gr-qc/0012036}{{\tt gr-qc/0012036}}].

\bibitem{Astefanesei:2009mc}
D.~Astefanesei, M.~J. Rodriguez, and S.~Theisen, {\it {Quasilocal equilibrium
  condition for black ring}},  {\em JHEP} {\bf 12} (2009) 040,
  [\href{http://xxx.lanl.gov/abs/0909.0008}{{\tt arXiv:0909.0008}}].

\bibitem{Harmark:2004ch}
T.~Harmark and N.~A. Obers, {\it General definition of gravitational tension},
  {\em JHEP} {\bf 05} (2004) 043,
  [\href{http://xxx.lanl.gov/abs/hep-th/0403103}{{\tt hep-th/0403103}}].

\bibitem{Kastor:2007wr}
D.~Kastor, S.~Ray, and J.~Traschen, {\it The first law for boosted
  {Kaluza--Klein} black holes},  {\em JHEP} {\bf 06} (2007) 026,
  [\href{http://xxx.lanl.gov/abs/arXiv:0704.0729 [hep-th]}{{\tt arXiv:0704.0729
  [hep-th]}}].

\bibitem{Carter:1989xk}
B.~Carter, {\it {Stability and characteristic propagation speeds in
  superconducting cosmic and other string models}},  {\em Phys. Lett.} {\bf
  B228} (1989) 466--470.

\bibitem{Carter:1997pb}
B.~Carter, {\it {Brane dynamics for treatment of cosmic strings and vortons}},
  \href{http://xxx.lanl.gov/abs/hep-th/9705172}{{\tt hep-th/9705172}}.

\bibitem{Emparan:2003sy}
R.~Emparan and R.~C. Myers, {\it Instability of ultra-spinning black holes},
  {\em JHEP} {\bf 09} (2003) 025,
  [\href{http://xxx.lanl.gov/abs/hep-th/0308056}{{\tt hep-th/0308056}}].

\bibitem{Cardoso:2006ks}
V.~Cardoso and O.~J.~C. Dias, {\it {Rayleigh-Plateau} and {Gregory-Laflamme}
  instabilities of black strings},  {\em Phys. Rev. Lett.} {\bf 96} (2006)
  181601, [\href{http://xxx.lanl.gov/abs/hep-th/0602017}{{\tt
  hep-th/0602017}}].

\bibitem{Caldarelli:2008mv}
M.~M. Caldarelli, O.~J.~C. Dias, R.~Emparan, and D.~Klemm, {\it {Black Holes as
  Lumps of Fluid}},  {\em JHEP} {\bf 04} (2009) 024,
  [\href{http://xxx.lanl.gov/abs/0811.2381}{{\tt 0811.2381}}].

\bibitem{Kol:2007rx}
B.~Kol and M.~Smolkin, {\it {Classical Effective Field Theory and Caged Black
  Holes}},  {\em Phys. Rev.} {\bf D77} (2008) 064033,
  [\href{http://xxx.lanl.gov/abs/0712.2822}{{\tt 0712.2822}}].

\bibitem{Galloway:2005mf}
G.~J. Galloway and R.~Schoen, {\it A generalization of {H}awking's black hole
  topology theorem to higher dimensions},  {\em Commun. Math. Phys.} {\bf 266}
  (2006) 571--576, [\href{http://xxx.lanl.gov/abs/gr-qc/0509107}{{\tt
  gr-qc/0509107}}].

\bibitem{Helfgott:2005jn}
C.~Helfgott, Y.~Oz, and Y.~Yanay, {\it On the topology of black hole event
  horizons in higher dimensions},  {\em JHEP} {\bf 02} (2006) 025,
  [\href{http://xxx.lanl.gov/abs/hep-th/0509013}{{\tt hep-th/0509013}}].

\bibitem{Hollands:2007aj}
S.~Hollands and S.~Yazadjiev, {\it {Uniqueness theorem for 5-dimensional black
  holes with two axial {K}illing fields}},  {\em Commun. Math. Phys.} {\bf 283}
  (2008) 749--768, [\href{http://xxx.lanl.gov/abs/0707.2775}{{\tt
  arXiv:0707.2775}}].

\bibitem{Evslin:2008gx}
J.~Evslin, {\it {Geometric Engineering 5d Black Holes with Rod Diagrams}},
  {\em JHEP} {\bf 09} (2008) 004,
  [\href{http://xxx.lanl.gov/abs/0806.3389}{{\tt arXiv:0806.3389}}].

\bibitem{Chen:2008fa}
Y.~Chen and E.~Teo, {\it {A rotating black lens solution in five dimensions}},
  {\em Phys. Rev.} {\bf D78} (2008) 064062,
  [\href{http://xxx.lanl.gov/abs/0808.0587}{{\tt arXiv:0808.0587}}].

\bibitem{Harmark:2009dh}
T.~Harmark, {\it {Domain Structure of Black Hole Space-Times}},  {\em Phys.
  Rev.} {\bf D80} (2009) 024019, [\href{http://xxx.lanl.gov/abs/0904.4246}{{\tt
  arXiv:0904.4246}}].

\bibitem{Kleihaus:2009wh}
B.~Kleihaus, J.~Kunz, and E.~Radu, {\it {$d\geq 5$ static black holes with
  $S^2\times S^{d-4}$ event horizon topology}},  {\em Phys. Lett.} {\bf B678}
  (2009) 301--307, [\href{http://xxx.lanl.gov/abs/0904.2723}{{\tt
  arXiv:0904.2723}}].

\bibitem{Emparan:2001wn}
R.~Emparan and H.~S. Reall, {\it A rotating black ring in five dimensions},
  {\em Phys. Rev. Lett.} {\bf 88} (2002) 101101,
  [\href{http://xxx.lanl.gov/abs/hep-th/0110260}{{\tt hep-th/0110260}}].

\bibitem{Elvang:2007rd}
H.~Elvang and P.~Figueras, {\it Black saturn},  {\em JHEP} {\bf 05} (2007) 050,
  [\href{http://xxx.lanl.gov/abs/hep-th/0701035}{{\tt hep-th/0701035}}].

\bibitem{Elvang:2007hg}
H.~Elvang, R.~Emparan, and P.~Figueras, {\it Phases of five-dimensional black
  holes},  {\em JHEP} {\bf 05} (2007) 056,
  [\href{http://xxx.lanl.gov/abs/hep-th/0702111}{{\tt hep-th/0702111}}].

\bibitem{Iguchi:2007is}
H.~Iguchi and T.~Mishima, {\it Black di-ring and infinite nonuniqueness},  {\em
  Phys. Rev.} {\bf D75} (2007) 064018,
  [\href{http://xxx.lanl.gov/abs/hep-th/0701043}{{\tt hep-th/0701043}}].

\bibitem{Evslin:2007fv}
J.~Evslin and C.~Krishnan, {\it {The Black Di-Ring: {A}n Inverse Scattering
  Construction}},  {\em Class. Quant. Grav.} {\bf 26} (2009) 125018,
  [\href{http://xxx.lanl.gov/abs/0706.1231}{{\tt 0706.1231}}].

\bibitem{Elvang:2007hs}
H.~Elvang and M.~J. Rodriguez, {\it {Bicycling Black Rings}},  {\em JHEP} {\bf
  04} (2008) 045, [\href{http://xxx.lanl.gov/abs/0712.2425}{{\tt 0712.2425}}].

\bibitem{Izumi:2007qx}
K.~Izumi, {\it {Orthogonal black di-ring solution}},  {\em Prog. Theor. Phys.}
  {\bf 119} (2008) 757--774, [\href{http://xxx.lanl.gov/abs/0712.0902}{{\tt
  0712.0902}}].

\bibitem{Harmark:2003yz}
T.~Harmark, {\it Small black holes on cylinders},  {\em Phys. Rev.} {\bf D69}
  (2004) 104015, [\href{http://xxx.lanl.gov/abs/hep-th/0310259}{{\tt
  hep-th/0310259}}].

\bibitem{Gorbonos:2004uc}
D.~Gorbonos and B.~Kol, {\it A dialogue of multipoles: Matched asymptotic
  expansion for caged black holes},  {\em JHEP} {\bf 06} (2004) 053,
  [\href{http://xxx.lanl.gov/abs/hep-th/0406002}{{\tt hep-th/0406002}}].

\bibitem{Elvang:2006dd}
H.~Elvang, R.~Emparan, and A.~Virmani, {\it Dynamics and stability of black
  rings},  {\em JHEP} {\bf 12} (2006) 074,
  [\href{http://xxx.lanl.gov/abs/hep-th/0608076}{{\tt hep-th/0608076}}].

\bibitem{Dias:2009iu}
O.~J.~C. Dias, P.~Figueras, R.~Monteiro, J.~E. Santos, and R.~Emparan, {\it
  {Instability and new phases of higher-dimensional rotating black holes}},
  {\em Phys. Rev.} {\bf D80} (2009) 111701,
  [\href{http://xxx.lanl.gov/abs/0907.2248}{{\tt arXiv:0907.2248}}].

\bibitem{Gubser:2001ac}
S.~S. Gubser, {\it On non-uniform black branes},  {\em Class. Quant. Grav.}
  {\bf 19} (2002) 4825--4844,
  [\href{http://xxx.lanl.gov/abs/hep-th/0110193}{{\tt hep-th/0110193}}].

\bibitem{Wiseman:2002zc}
T.~Wiseman, {\it Static axisymmetric vacuum solutions and non-uniform black
  strings},  {\em Class. Quant. Grav.} {\bf 20} (2003) 1137--1176,
  [\href{http://xxx.lanl.gov/abs/hep-th/0209051}{{\tt hep-th/0209051}}].

\end{thebibliography}

\providecommand{\href}[2]{#2}\begingroup\raggedright\endgroup

\end{document}